\UseRawInputEncoding
\documentclass[%
 reprint,
 superscriptaddress,
 amsmath,amssymb,
 aps,
 prl % email on first or before references
]{revtex4-2}

\usepackage[pagewise,modulo]{lineno}
\usepackage{xcolor}
\usepackage{graphicx}% Include figure files
\usepackage{dcolumn}% Align table columns on decimal point
\usepackage{bm}% bold math
\usepackage[hidelinks,colorlinks=true,linkcolor=blue,urlcolor=blue,citecolor=blue]{hyperref}
\makeatletter
\newcommand{\mycaption}[2][]{%
  \begingroup%
  \renewcommand{\figurename}{\textbf{Figure}}
  \renewcommand{\@caption@fignum@sep}{ \textbf{$\vert$} }%
  \renewcommand{\fnum@figure}{{\normalfont\bfseries \figurename~\thefigure}}
  \caption[#1]{#2}%
  \endgroup%
}
\makeatother

\makeatletter
\def\ignorecitefornumbering#1{%
     \begingroup
         \@fileswfalse
         #1%                     % do \cite comand
    \endgroup
}
\makeatother

\begin{document}

%\preprint{APS/123-QED}

\title{Strain-tunable Berry curvature in quasi-two-dimensional chromium telluride}
\author{Hang Chi}    
    \email{chihang@mit.edu}
    \affiliation{Francis Bitter Magnet Laboratory, Plasma Science and Fusion Center, Massachusetts Institute of Technology, Cambridge, Massachusetts 02139, USA}
    \affiliation{U.S. Army CCDC Army Research Laboratory, Adelphi, Maryland 20783, USA}
\author{Yunbo Ou}    
    \email{ybou@mit.edu}
    \affiliation{Francis Bitter Magnet Laboratory, Plasma Science and Fusion Center, Massachusetts Institute of Technology, Cambridge, Massachusetts 02139, USA}
\author{Tim B. Eldred}    
    \affiliation{Department of Materials Science and Engineering, North Carolina State University, Raleigh, North Carolina 27695, USA}
\author{Wenpei Gao}    
    \affiliation{Department of Materials Science and Engineering, North Carolina State University, Raleigh, North Carolina 27695, USA}
\author{Sohee Kwon}    
    \affiliation{Department of Electrical and Computer Engineering, 
University of California, Riverside, California 92521, USA}
\author{Joseph Murray}    
    \affiliation{Department of Physics, University of Maryland, College Park, Maryland 20742, USA}
\author{Michael Dreyer}    
    \affiliation{Department of Physics, University of Maryland, College Park, Maryland 20742, USA}
\author{Robert E. Butera}    
    \affiliation{Laboratory for Physical Sciences, College Park, Maryland 20740, USA}
\author{Alexandre C. Foucher}    
    \affiliation{Department of Materials Science and Engineering, Massachusetts Institute of Technology, Cambridge, Massachusetts 02139, USA}
\author{Haile Ambaye}    
    \affiliation{Neutron Scattering Division, Neutron Sciences Directorate, Oak Ridge National Laboratory, Oak Ridge, Tennessee 37831, USA}
\author{Jong Keum}    
    \affiliation{Neutron Scattering Division, Neutron Sciences Directorate, Oak Ridge National Laboratory, Oak Ridge, Tennessee 37831, USA}
    \affiliation{Center for Nanophase Materials Sciences, Physical Science Directorate, Oak Ridge National Laboratory, Oak Ridge, Tennessee 37831, USA}
\author{Alice T. Greenberg}    
    \affiliation{U.S. Army CCDC Army Research Laboratory, Adelphi, Maryland 20783, USA}
\author{Yuhang Liu}    
    \affiliation{Department of Electrical and Computer Engineering, 
University of California, Riverside, California 92521, USA}
\author{Mahesh R. Neupane}    
    \affiliation{U.S. Army CCDC Army Research Laboratory, Adelphi, Maryland 20783, USA}
    \affiliation{Department of Electrical and Computer Engineering, 
University of California, Riverside, California 92521, USA}
\author{George J. de Coster}    
    \affiliation{U.S. Army CCDC Army Research Laboratory, Adelphi, Maryland 20783, USA}
\author{Owen A. Vail}    
    \affiliation{U.S. Army CCDC Army Research Laboratory, Adelphi, Maryland 20783, USA}
\author{Patrick J. Taylor}    
    \affiliation{U.S. Army CCDC Army Research Laboratory, Adelphi, Maryland 20783, USA}
\author{Patrick A. Folkes}    
    \affiliation{U.S. Army CCDC Army Research Laboratory, Adelphi, Maryland 20783, USA}
\author{Charles Rong}    
    \affiliation{U.S. Army CCDC Army Research Laboratory, Adelphi, Maryland 20783, USA}
\author{Gen Yin}    
    \affiliation{Department of Physics, Georgetown University, Washington, District of Columbia 20057, USA}
\author{Roger K. Lake}    
    \affiliation{Department of Electrical and Computer Engineering, 
University of California, Riverside, California 92521, USA}
\author{Frances M. Ross}    
    \affiliation{Department of Materials Science and Engineering, Massachusetts Institute of Technology, Cambridge, Massachusetts 02139, USA}
\author{Valeria Lauter}    
    \affiliation{Neutron Scattering Division, Neutron Sciences Directorate, Oak Ridge National Laboratory, Oak Ridge, Tennessee 37831, USA}
\author{Don Heiman}    
    \affiliation{Francis Bitter Magnet Laboratory, Plasma Science and Fusion Center, Massachusetts Institute of Technology, Cambridge, Massachusetts 02139, USA}
    \affiliation{Department of Physics, Northeastern University, Boston, Massachusetts 02115, USA}
\author{Jagadeesh S. Moodera}    
    \email{moodera@mit.edu}
    \affiliation{Francis Bitter Magnet Laboratory, Plasma Science and Fusion Center, Massachusetts Institute of Technology, Cambridge, Massachusetts 02139, USA}
    \affiliation{Department of Physics, Massachusetts Institute of Technology, Cambridge, Massachusetts 02139, USA}

\date{\today}
%\date{July 6, 2022}
%%% Abstract %%%

\begin{abstract}
Magnetic transition metal chalcogenides form an emerging platform for exploring spin-orbit driven Berry phase phenomena owing to the nontrivial interplay between topology and magnetism. Here we show that the anomalous Hall effect in pristine Cr$_2$Te$_3$ thin films manifests a unique temperature-dependent sign reversal at nonzero magnetization, resulting from the momentum-space Berry curvature as established by first-principles simulations. The sign change is strain tunable, enabled by the sharp and well-defined substrate/film interface in the quasi-two-dimensional Cr$_2$Te$_3$ epitaxial films, revealed by scanning transmission electron microscopy and depth-sensitive polarized neutron reflectometry. This Berry phase effect further introduces hump-shaped Hall peaks in pristine Cr$_2$Te$_3$ near the coercive field during the magnetization switching process, owing to the presence of strain-modulated magnetic domains. The versatile interface tunability of Berry curvature in Cr$_2$Te$_3$ thin films offers new opportunities for topological electronics. 
\end{abstract}

\maketitle

\begin{figure*}
\includegraphics{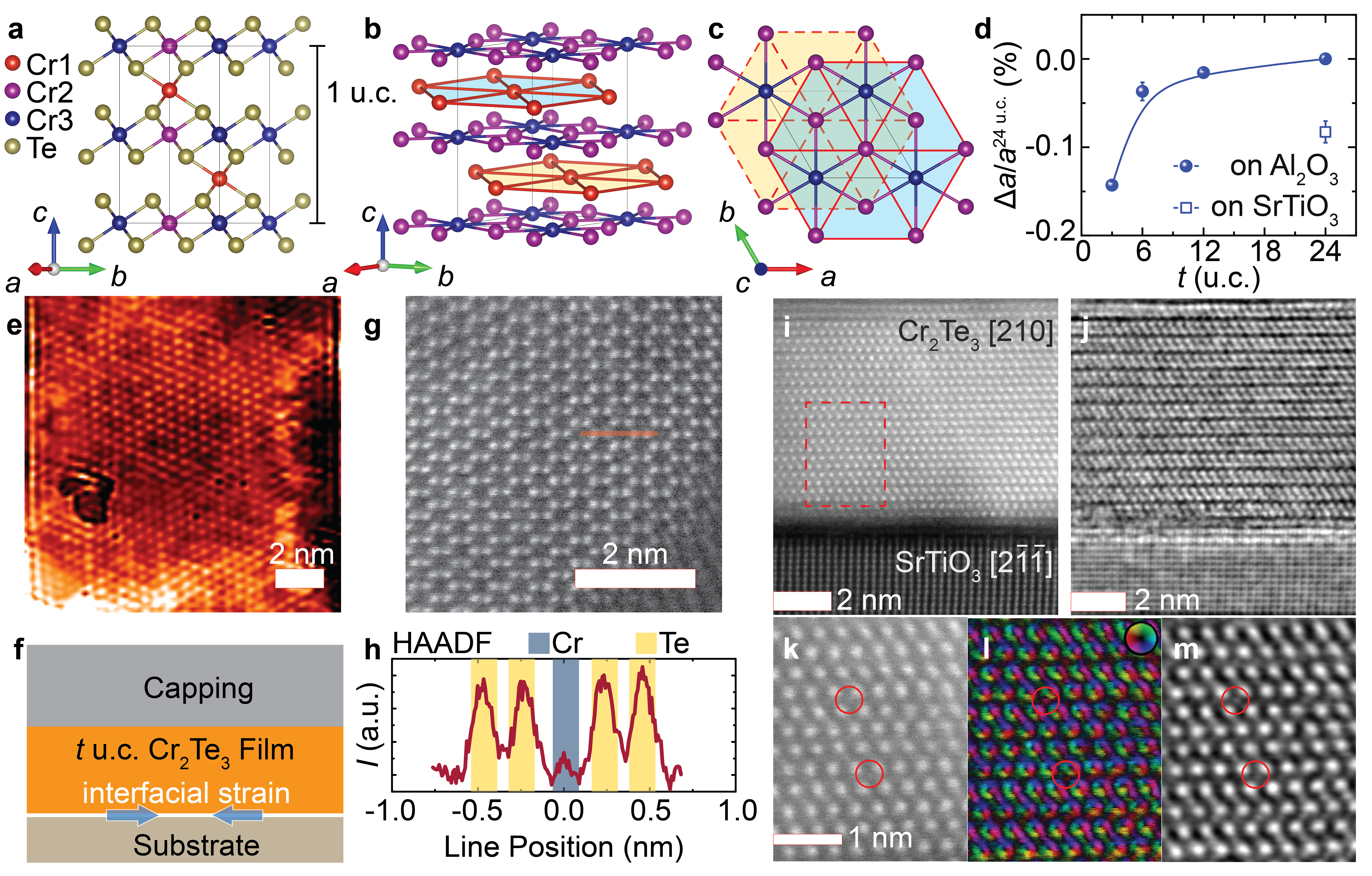}
\mycaption{\label{fig:fig1}\textbf{Crystal structure of Cr$_2$Te$_3$ thin films.}
\textbf{a}, Atomistic structure of Cr$_2$Te$_3$ viewed along the crystallographic [210] direction. \textbf{b}, Among the three Cr species, Cr1 (red) form sparse honeycombs that are stacked between those of Cr2/Cr3 (purple/blue) with six-fold in-plane symmetry (\textbf{c}). \textbf{d}, Enhanced in-plane compressive strain at reduced thickness $t$, quantified by the relative change of the $a$ lattice parameter via XRD for Cr$_2$Te$_3$ grown on Al$_2$O$_3$(0001) (solid) or SrTiO$_3$(111) (open). \textbf{f}, Schematic of the film stacks, where the interfacial strain plays a pivotal role in inducing extraordinary magnetic and transport phenomena. Atomically resolved STM morphology of a $13 \times 13$ nm$^2$ surface after removing Se capping (\textbf{e}) and planar HAADF STEM image (\textbf{g}) of Cr$_2$Te$_3$ confirm the honeycomb-like Te lattice, where the HAADF intensity line scan (\textbf{h}) reveals the Cr sites. \textbf{i}-\textbf{m}, Cross-sectional images of Cr$_2$Te$_3$ films grown on SrTiO$_3$(111). The HAADF (\textbf{i}) and iDPC (\textbf{j}) imaging along the [210] zone axis of Cr$_2$Te$_3$ illustrates the dominating Te-Cr2/Cr3-Te layers. The enlarged view (dashed box region in \textbf{i}) of HAADF (\textbf{k}), DPC (\textbf{l}), and iDPC (\textbf{m}) images identify the random distribution of the interlayer Cr1 (circles), which deviates from the ideal Cr$_2$Te$_3$ structure with full occupancy. The color wheel in the DPC image indicates the projected electric field direction.
}
\end{figure*}

In recent years, a variety of novel two-dimensional (2D) van der Waals magnets have been discovered, founding the active field of 2D magnetism \cite{burch-2018-review-nature}.\ Among these prospective compounds, binary chromium tellurides Cr$_{1-\delta}$Te \cite{Ipser-1983-JLCM, McGuire-2017-PRB-0.333, Zhang-2021-NC-0.5, Liu-2018-PRB-0.625, Wen-2020-NL-0.667, Cao-2019-PRM-0.75x, Chua-2021-AM-0.75} are attractive owing to their rich magnetic properties, as well as inherent chemical and structural compatibility when forming heterostructures \cite{bib3} with other topological systems, such as tetradymite-type topological insulators \cite{bib4} or chalcogenide-based Dirac/Weyl semimetals \cite{armitage-2018-rmp}. Furthermore, the broken time-reversal symmetry and the strong spin-orbit coupling (SOC) offer unique opportunities for the interplay between spin configurations and reciprocal-space topology \cite{Jiang-2020-NM,Bernevig-2022-review-nature,Chi-QAH}. In this regard, ferromagnetic Cr$_2$Te$_3$ with strong perpendicular magnetic anisotropy (PMA) is an intriguing platform to host non-trivial topological physics, particularly for the high-quality thin films grown by molecular beam epitaxy (MBE) \cite{bib6,bib7}. 

An important consequence of the band topology in Cr$_2$Te$_3$ is the Berry curvature \cite{bib8,bib9} underlying the anomalous Hall effect (AHE) \cite{bib10}. The intrinsic AHE is topological in nature and a hallmark of itinerant ferromagnets, which has also been observed in more exotic systems even without a net magnetization, such as spin liquids \cite{bib11}, antiferromagnets \cite{bib12} and Weyl semimetals \cite{bib13}. When SOC coexists with long-range magnetic order, the Berry curvature can be significantly influenced near avoided band crossings, rendering the system an incredibly rich playground combining topology and magnetism \cite{bib14, bib15}. 

Here, we report the unique magnetotransport signatures of high-quality quasi-2D Cr$_2$Te$_3$ MBE grown thin films governed by non-trivial band topologies. Via synergetic structural, magnetic and transport measurements, together with first-principles simulations, we have uncovered novel Berry-curvature-induced magnetism featuring an extraordinary sign reversal of the AHE, as we modulate the temperature and the strain for the thin films containing 3 to 24 unit cells (u.c.) on Al$_2$O$_3$(0001) or SrTiO$_3$(111) substrates. Moreover, a hump-shaped Hall feature emerges, most likely due to the presence of multiple magnetic domains under different levels of interfacial strain. This work identifies pristine ferromagnetic Cr$_2$Te$_3$ thin films as a fascinating platform for further engineering topological effects given their nontrivial Berry curvature physics.

%\noindent\textbf{Results} 

\noindent\textbf{Atomic structure, interfaces and strain.} 

The crystalline structure of Cr$_2$Te$_3$ thin films is described first, followed by the development of strain at the substrate/film interface by the epitaxy. Bulk Cr$_2$Te$_3$ crystalizes in the structure with space group $P\bar{3}1c$ ($D_{3d}^2$, No. 163), as shown in Figs.~\ref{fig:fig1}a-c, where each unit cell contains four vertically stacked hexagonal layers of Cr \cite{bib16}. There are three symmetrically unique sites for Cr, labeled Cr1, Cr2 and Cr3, respectively: The Cr1 atoms are sparsely arranged in a weakly antiferromagnetic sublattice \cite{bib17}, while the Cr2/Cr3 atoms form ferromagnetic layers similar to those in CrTe$_2$ \cite{bib18}. Since the Cr1 sites are often only partially filled (Figs.~\ref{fig:fig1}i-m), Cr$_2$Te$_3$ behaves essentially as a quasi-2D magnet \cite{bib20,bib19,bib21}. This quasi-2D nature of Cr$_2$Te$_3$ allows for high-quality, layer-by-layer epitaxial growth of $c$-oriented films on a variety of substrates. The hexagonal $c$ axis is the easy magnetic axis, leading to PMA for the films.

The six-fold in-plane (IP) symmetry is seen in the honeycombs visualized by atomic resolution scanning tunneling microscopy (STM, Fig.~\ref{fig:fig1}e) and scanning transmission electron microscopy (STEM, Fig.~\ref{fig:fig1}g) high-angle annular dark-field (HAADF) imaging, as well as in the reflection high-energy electron diffraction (RHEED, Supplementary Fig.~1) and X-ray diffraction (XRD, Supplementary Fig.~2) patterns. The sharp substrate/film interface is confirmed by the cross sectional HAADF (Fig.~\ref{fig:fig1}i) and the corresponding integrated differential phase contrast (iDPC, Fig.~\ref{fig:fig1}j) images. The intrinsic random distribution of Cr atoms on the Cr1 sites is resolved in the enlarged view of the atoms in Figs.~\ref{fig:fig1}k-m, shown overlaid with red circles, while the overall chemical composition of the thin film is uniform within the resolution of energy dispersive X-ray spectroscopy (EDS, see Supplementary Fig. 3). 

\begin{figure*}
\includegraphics{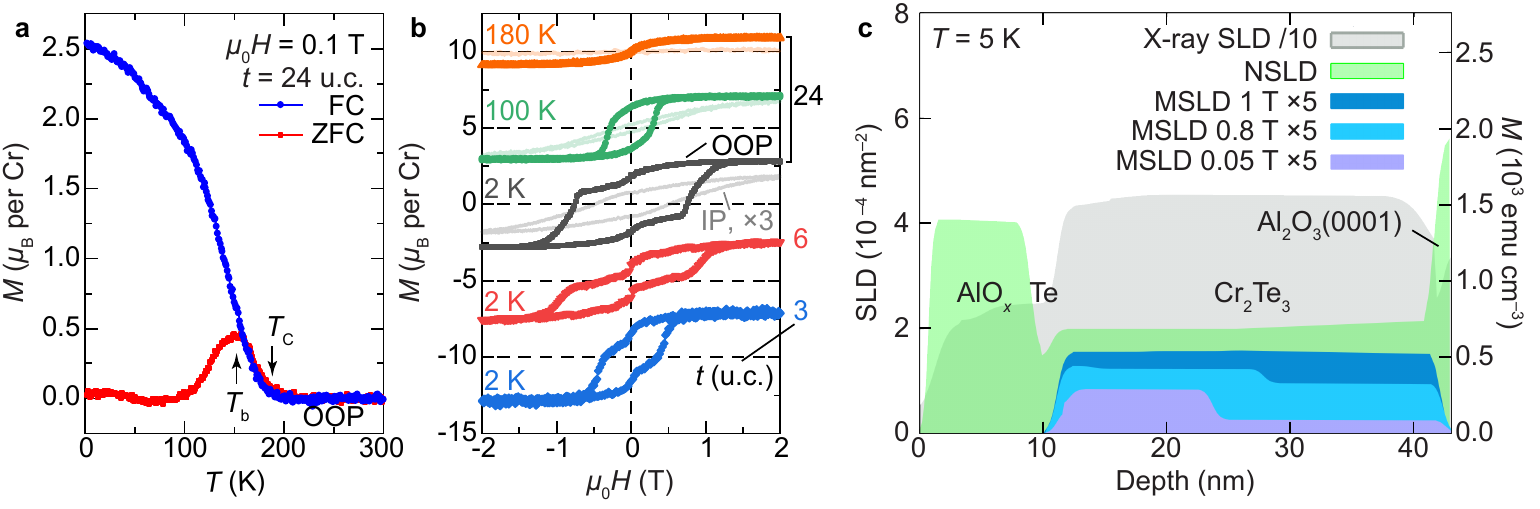}
\mycaption{\label{fig:fig2}\textbf{Magnetic properties of Cr$_2$Te$_3$ thin films.} \textbf{a}, Temperature dependence of the magnetization $M$ of a typical 24~u.c.\ Cr$_2$Te$_3$ film under the zero-field-cool (ZFC) and field-cool (FC) conditions with an out-of-plane (OOP) external magnetic field $\mu_{0}H$ = 0.1 T. The Curie ($T_{\textrm{C}}$) and blocking ($T_{\textrm{b}}$) temperatures are labeled by the arrows. \textbf{b}, Field dependence of $M$ under OOP and in-plane (IP) configurations for $t$ = 24 u.c.\ at selected temperatures (top three, black, green and orange) and OOP $M(H)$ for $t$ = 6 u.c. and 3 u.c.\ at 2 K (bottom two, red and blue). For clarity, the curves are vertically shifted and the IP data are magnified by a factor of 3. \textbf{c}, Depth profiles of PNR nuclear (NSLD), magnetic (MSLD, at IP fields of 1 T, 0.8 T and 0.05 T, respectively) and X-ray scattering length densities (SLD) of 24 u.c.\ Cr$_2$Te$_3$ on Al$_2$O$_3$(0001) with Te/AlO$_x$ capping.}
\end{figure*}

Figure~\ref{fig:fig1}f illustrates the basic sample architecture, where the strain in the Cr$_2$Te$_3$ thin films is governed by the interface with the substrate. Upon reducing the thickness ($t$), films grown on Al$_2$O$_3(0001)$ can develop an IP compressive strain up to $-0.15$\%, as determined by XRD and summarized in Fig.~\ref{fig:fig1}d. A higher strain level can be sustained using SrTiO$_3$(111) substrates. Such control of strain is well suited for exploring interface-sensitive properties in Cr$_2$Te$_3$ thin films.

\noindent\textbf{Magnetism, domains and PNR.}  

The magnetic properties of Cr$_2$Te$_3$ thin films with selected thicknesses were assessed using vibrating sample magnetometry (VSM). Figure~\ref{fig:fig2}a shows the temperature dependence of the magnetization $M(T)$ for a $t$ = 24 u.c.\ film on Al$_2$O$_3$(0001) substrate with an out-of-plane (OOP) applied magnetic field $\mu_{0}H$ = 0.1 T. Under the field-cool (FC) condition, $M(T)$ rises below the Curie temperature $T\textsubscript{C}$ $\sim$ 180 K, reaching $M \sim$ 2.50 $\mu_\textsubscript{B}$ (Bohr magneton) per Cr at 2 K in the 0.1 T field. The zero-field-cool (ZFC) scan on the other hand deviates from the FC curve below the blocking temperature $T\textsubscript{b}$, signaling the freezing out of domains in random direction in the absence of an aligning field $H$. 

As illustrated in Fig.~\ref{fig:fig2}b, Cr$_2$Te$_3$ favors PMA with coercive field $\mu_{0}H_{\textrm{c}}$ = 0.76 T and saturation magnetization $M\textsubscript{s} \sim$ 2.83  $\mu_{\textrm{B}}$ per Cr at 2 K for $t$ = 24 u.c., whereas the IP measurements have weaker ferromagnetic hysteresis loops. The low-$T$ zero-field kink in the OOP $M(H)$ becomes more prominent at reduced $t$ (for additional data on $t$ = 6 u.c., see Supplementary Fig.~4), indicating the presence of interfacial strain-induced multiple magnetic domains with continuously varying spin canting \cite{bib22, bib23, bib24}. 

Depth-sensitive polarized neutron reflectometry (PNR) measurements, responsive to the IP magnetization, were carried out at chosen $T$ and $H$ on samples with $t$ = 24 and 6 u.c., in order to uncover the impact of interfacial strain. The PNR spin asymmetry ratio SA = $(R^{+}-R^{-})/(R^{+}+R^{-})$, measured as a function of the wave vector transfer $Q = 4\pi\sin(\theta)/\lambda$ with $R^+$ and $R^-$ being the reflectivity for the neutron spin parallel ($+$) or antiparallel ($-$) to the external field, evidently confirms the magnetization (Supplementary Fig.~5). By simultaneously refining PNR and X-ray reflectivity (XRR, Supplementary Fig.~5) data, the depth profiles of nuclear (NSLD) and magnetic (MSLD) scattering length densities (SLD) at $\mu_{0}H$ = 1 T, 0.8 T and 0.05 T for $t$ = 24 u.c. were obtained and shown in Fig.~\ref{fig:fig2}c. The uniform MSLD profile at the IP saturation field $\mu_{0}H$ = 1 T attests to the high quality of the magnetic Cr$_2$Te$_3$ layer with well-defined interfaces of 0.5 nm. Remarkably, at a reduced IP field $\mu_{0}H$ = 0.8 T and 0.05 T, $M$ develops a non-uniform depth-dependent magnetization profile with two components, revealing a lower (higher) value close to (away from) the substrate. Since PNR is sensitive to IP magnetization, these results collectively suggest that more pronounced strain at the interface leads to a higher OOP magnetic anisotropy and hence a lower measured IP MSLD. This scenario is further substantiated by the lower $M$ observed for $t$ = 6 u.c. with stronger strain measured at 5 K and 60 K under 1 T IP magnetic field (Supplementary Fig.~4d). The salient structural and magnetic features pave the way for an in-depth investigation of the magneto-transport responses in Cr$_2$Te$_3$ thin films.\\ 

\begin{figure*}
\includegraphics{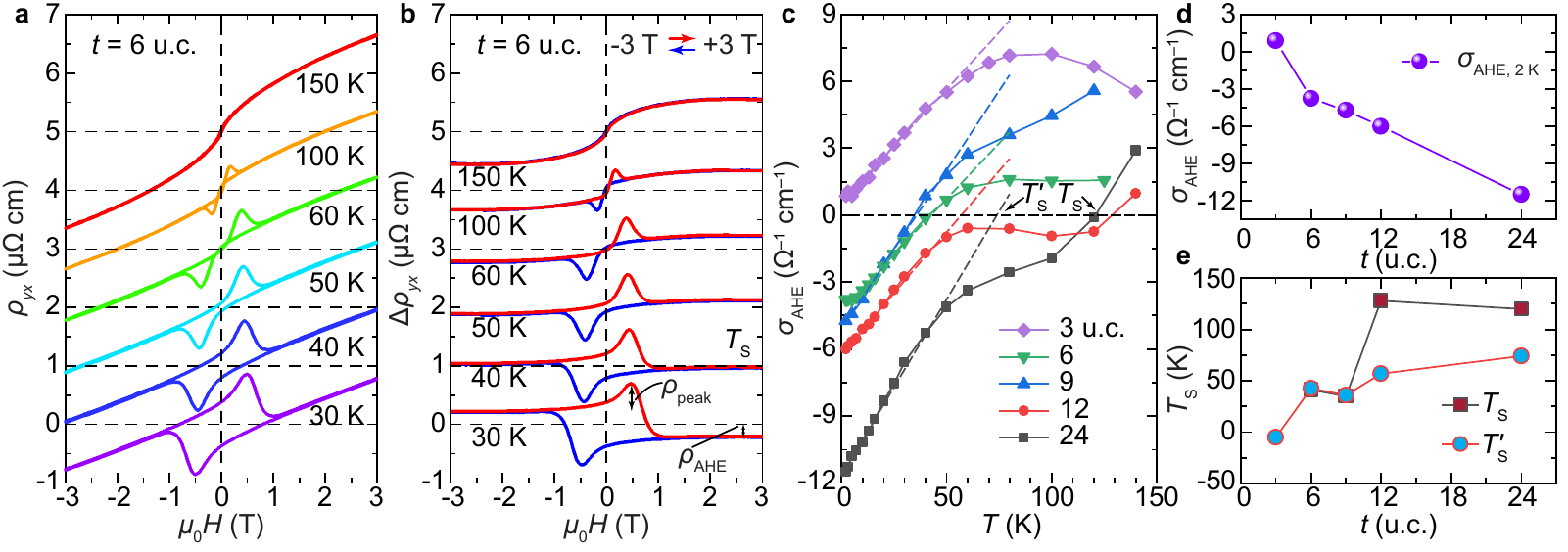}
\mycaption{\label{fig:fig3}\textbf{The unconventional Hall effects in Cr$_2$Te$_3$ thin films.} \textbf{a}, Magnetic field dependence of the Hall resistivity $\rho_{yx}(H)$ at selected temperatures of 6 u.c.\ Cr$_2$Te$_3$ on Al$_2$O$_3$(0001). \textbf{b}, Hall traces $\Delta\rho_{yx}$ after removing the high-field ordinary Hall backgrounds. At $T\textsubscript{S}$ $\sim$ 40 K, a sign change occurs in the anomalous Hall resistivity $\rho\textsubscript{AHE}$, defined as the value of $\Delta\rho_{yx}$ when the system is fully magnetized under a positive $H$. Apart from the AHE hysteresis loop, additional hump-shaped features develop. \textbf{c}, Temperature dependence of the anomalous Hall conductivity $\sigma_{\textrm{AHE}}$ for $t$ = 3 -- 24 u.c. (symbols, where solid lines are guide for the eye and dashed lines are linear fit to the low $T$ data). \textbf{d}-\textbf{e}, Thickness dependence of $\sigma_{\textrm{AHE}}$ at 2 K (\textbf{d}), AHE sign reversal temperature $T\textsubscript{S}$ (\textbf{e}) and $T_{\textrm{S}}^{'}$, the $T$-intercept of the linear AHE component at low $T$.}
\end{figure*}

\noindent\textbf{Strain-tunable AHE and sign reversal.}

The unusual Hall effects are the most outstanding properties of the Cr$_2$Te$_3$ thin films. The development of long-range magnetic ordering is manifested in the AHE-induced hysteresis in the Hall resistivity 
\begin{equation}
\rho_{yx}(H)=R\textsubscript{H}+R\textsubscript{S}M,
\label{eq:rhoyx}
\end{equation}
in Fig.~\ref{fig:fig3}a (for more details on the transport parameters, see Supplementary Fig.~6). Here $R_{\textrm{H}}$ characterizes the linear-in-$H$ ordinary Hall effect (OHE) that dominates at high $H$ and $R_{\textrm{S}}$ is the AHE coefficient denoting contribution from the underlying magnetic order. 

By removing the linear OHE background in Fig.~\ref{fig:fig3}a, we now turn to the rich $T$ and $H$ dependences of the Hall traces $\Delta\rho_{yx}(H)$ and the unconventional AHE in the ferromagnetic regime in Fig.~\ref{fig:fig3}b. For $t$ = 6 u.c., at $T \leqslant$ 30 K, when fully magnetized under a positive $H$, the system produces a negative AHE signal $\rho\textsubscript{AHE}$, i.e., $\Delta\rho_{yx}(H)$ loops around the origin in the opposite direction of that for the $M(H)$ hysteresis (see Supplementary Fig.~4b). The $T$ dependence of the corresponding anomalous Hall conductivity $\sigma_{\textrm{AHE}}=\rho_{\textrm{AHE}}/(\rho^2_{\textrm{AHE}}+\rho^2_{xx})$, with $\rho_{xx}$ being the longitudinal electrical resistivity, is summarized in Fig.~\ref{fig:fig3}c. Upon rising $T$, $\rho_{\textrm{AHE}}$ surprisingly changes sign at a transition temperature $T\textsubscript{S}$ $\sim$ 40 K for $t$ = 6 u.c.. Note that the sign change signifies a compensation point at $T_{\textrm{S}}$ where $\rho_{\textrm{AHE}}$ or the anomalous Hall conductivity $\sigma_{\textrm{AHE}}$ traverses through zero while $M$ remains finite (see Supplementary Fig. 7b). This is a highly unusual behavior, although rather similar to the transport anomaly present in SrRuO$_3$ as a result of the nontrivial band topology \cite{bib14}. Furthermore, this unique sign reversal behavior of the AHE is sensitively governed by the interfacial strain (Fig.~\ref{fig:fig3}d). As evident in Fig.~\ref{fig:fig3}e, $T_{\textrm{S}}$ largely decreases upon increasing compressive strain at reduced $t$ (Fig.~\ref{fig:fig1}d). At $t$ = 3 u.c., the strain is found to be sufficient in driving $\sigma_{\textrm{AHE}} > 0$ in the ground state, leading to the absence of a temperature-induced sign flipping at finite $T$.

\begin{figure*}
\includegraphics{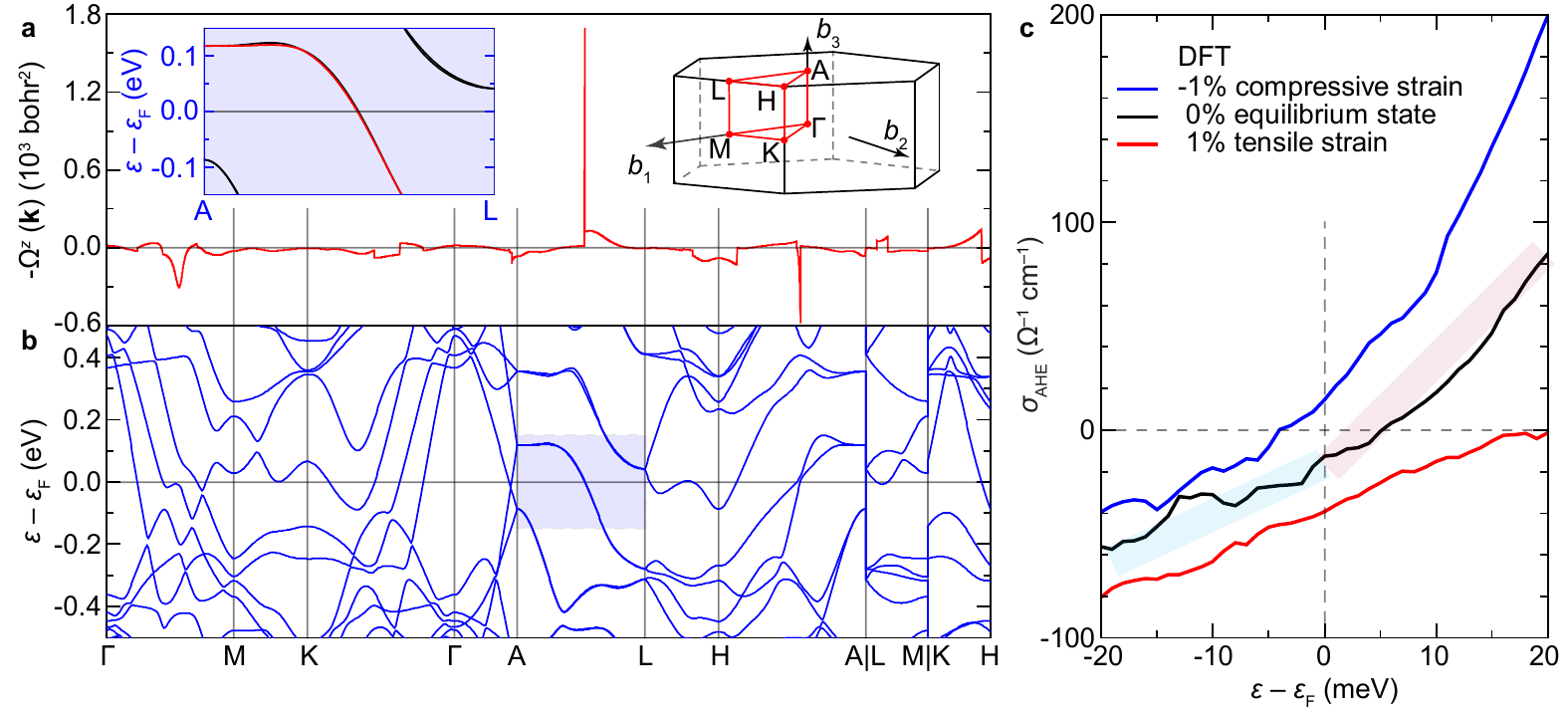}
\mycaption{\label{fig:fig4}\textbf{Berry curvature and anomalous Hall conductivity in Cr$_2$Te$_3$.} \textbf{a}-\textbf{b}, Calculated Berry curvature $\Omega^z(\mathbf{k})$ (\textbf{a}) along the high symmetry $k$-paths in the Brillouin zone (right inset in \textbf{a}) and the corresponding electronic band structure (\textbf{b}). Left inset in \textbf{a}, nearly degenerate SOC anti-crossing bands contributing to the sharp peak in $\Omega^z(\mathbf{k})$ along A-L. \textbf{c}, Anomalous Hall conductivity $\sigma_{\textrm{AHE}}$ near the Fermi level $\varepsilon_{\textrm{F}}$, in equilibrium state (black), under compressive (blue) or tensile (red) strain conditions, respectively. The shades in \textbf{c} are guide for the eye showing the slight asymmetry of the energy dependence of $\sigma_{\textrm{AHE}}$ above and below $\varepsilon_{\textrm{F}}$ which at finite $T$ may lead to a sign reversal in $\sigma_{\textrm{AHE}}$ owing to thermal broadening.}
\end{figure*}

To elucidate the physical origin of the AHE sign reversal of Cr$_2$Te$_3$, we examined the Berry curvature $\Omega^z(\mathbf{k})=\sum_{n}{f_n\Omega_n^z(\mathbf{k})}$ (Fig.~\ref{fig:fig4}a, summed over the occupied bands with $f_n$ the equilibrium Fermi-Dirac distribution function) based on the electronic band structure (Fig.~\ref{fig:fig4}b) obtained using density functional theory (DFT). As exemplified by the left inset of Fig.~\ref{fig:fig4}a, a significant spike feature develops in $\Omega^z(\mathbf{k})$, originating from the nearly degenerate SOC anti-crossing bands along the A-L $k$-path. The intrinsic AHE conductivity is evaluated by integrating over the Brillouin zone (BZ)
\begin{equation}
\sigma_{\textrm{AHE}}=-\frac{e^2}{\hbar}\int_{\textrm{BZ}}{\frac{d^3k}{{(2\pi)}^3}\Omega^z(\mathbf{k})},
\label{eq:sigmaxy}
\end{equation}
where $e$ is the electron charge and $\hbar$ is the reduced Planck's constant. As shown in Fig.~\ref{fig:fig4}c, the calculated $\sigma_{\textrm{AHE}}$ = $-12.7$ $\Omega^{-1}$~cm$^{-1}$ at the Fermi level $\varepsilon_{\textrm{F}}$ for Cr$_2$Te$_3$ under equilibrium state (the black curve in Fig.~\ref{fig:fig4}c, see also Supplementary Fig.~8 for the convergence test under different $k$-mesh), which is in excellent agreement with the experimental value of $-11.5$ $\Omega^{-1}$~cm$^{-1}$ for $t$ = 24 u.c.. It attests to the dominance of the intrinsic Berry phase mechanism, rather than the extrinsic side jump or skew scattering \cite{bib10}, as the primary origin of the observed AHE in Cr$_2$Te$_3$. The calculation also reveals a sensitive energy dependence of $\sigma_{\textrm{AHE}}$ -- not only the magnitude but also the sign changes near $\varepsilon_{\textrm{F}}$. At finite $T$, due to the thermal broadening in $f_n$, the slight asymmetry of $\sigma_{\textrm{AHE}}$ above and below $\varepsilon_{\textrm{F}}$, naturally explains the experimentally observed AHE sign anomaly. Modeling of strained cases in Fig.~\ref{fig:fig4}c further reveals that $\sigma_{\textrm{AHE}}$ at $\varepsilon_{\textrm{F}}$ changes sign under $-1$\% compressive strain, substantiating that Berry physics underlies the observed strain-driven AHE sign reversal in Fig.~\ref{fig:fig3}d. This unique capability of achieving zero $\sigma_{\textrm{AHE}}$ or $\rho_{\textrm{AHE}}$ while maintaining nonzero $M$ in Cr$_2$Te$_3$ thin films, deviating from the classic Eq.~(\ref{eq:rhoyx}), offers direct insight into the intrinsic AHE solely owing to the Berry curvature \cite{bib10, wideBerry}.\\

\noindent\textbf{Hump-shaped Hall peaks at the coercive field.}

\begin{figure}
\includegraphics{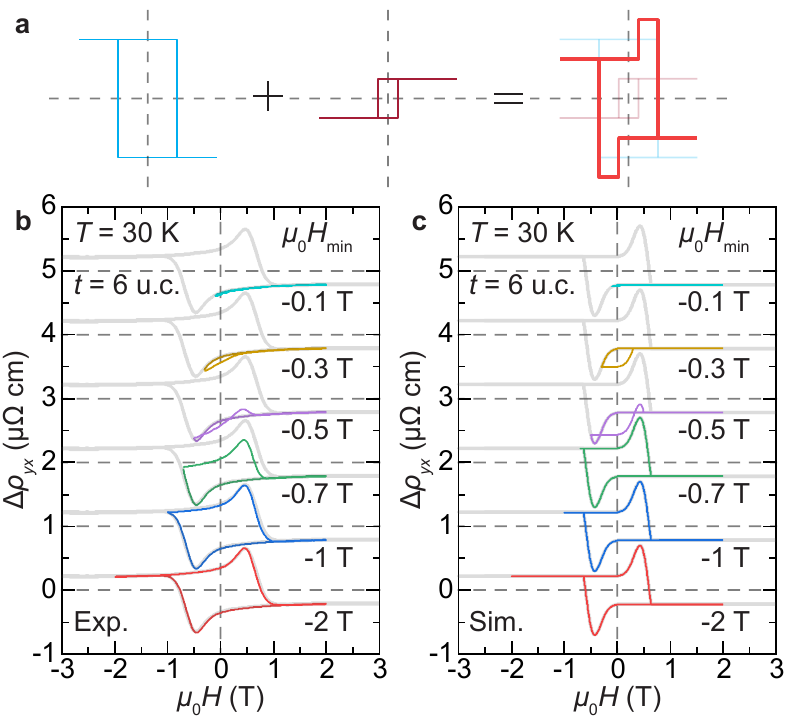}
\mycaption{\label{fig:fig5}\textbf{Characteristics of hump-shaped Hall peaks.} \textbf{a}, Simplified superposition of two AHE components with opposite sign and different coercive fields. \textbf{b}-\textbf{c}, Minor loop experiments on a 6 u.c.\ Cr$_2$Te$_3$ on Al$_2$O$_3$(0001) at $T$ = 30 K, first fully magnetized at $\mu_{0}H$ = +3 T (complete loop shown in grey as guide for the eye) and then swept back and forth between +2 T and selected $\mu_{0}H_{\textrm{min}}$. The experimental minor loops in \textbf{b} are qualitatively reproduced in \textbf{c} using simulations that underscore the significance of strain-driven multidomain features and the sign reversal in $\rho_{\textrm{AHE}}$.}
\end{figure}

Figure~\ref{fig:fig3}b also shows additional hump-shaped peaks on top of the otherwise square AHE hysteresis loop. The peaks are centered at the characteristic fields $H\textsubscript{peak}$ that tracks well with the coercive fields $H_{\textrm{c}}$ determined from the magnetic measurements (Supplementary Fig.~7). These hump-shaped Hall peaks in our pristine Cr$_2$Te$_3$ are related to the presence of strain-modulated magnetic multidomain structures with opposite signs of AHE (Fig.~\ref{fig:fig5}a), rather than the skyrmion-induced topological Hall effect as postulated in various related phases and heterostructures \cite{bib28, bib30, bib29, Zhang-crte2-bi2te3, Jeon-cr2te3-cr2se3, Ou-crte2-zrte2, Kimbell-2022-CM}. To better understand the mechanism(s) underlying the Hall peaks observed in $\Delta\rho_{yx}(H)$, minor loop experiments were carried out for $t$ = 6 u.c.\ at $T$ = 30 K and are shown in Fig.~\ref{fig:fig5}b. For each scan, the loop starts from a well-defined initial state that is fully magnetized under a positive $H$, which is then swept towards a negative $H\textsubscript{min}$ around $-H_{\textrm{peak}}$ and scanned back to the initial positive $H$. The minor loops are hysteretic, where the emergence of the Hall peak with positive $H$ depends on whether $H\textsubscript{min}$ surpasses $-H_{\textrm{peak}}$. The two-component origin of the Hall anomaly peaks can be quite well explained by a distribution of domains having $T$-dependent $H_{\textrm{c}}$ using \cite{bib27}
\begin{equation}
\Delta\rho_{yx}(H) = \int_{0}^{\infty} {\widetilde{\rho}}_{\textrm{AHE}}(T^\prime)\{2\textrm{H}_{\textrm{Heav}}[H-\widetilde{H}_{\textrm{c}}(T^\prime)]-1\}G(T^\prime) \,dT^\prime.
\label{eq:rhoyx-fit}
\end{equation}
Here $\widetilde{\rho}\textsubscript{AHE}(T^\prime)$ and $\widetilde{H}\textsubscript{c}(T^\prime)$ are functionals based on experimental $\rho\textsubscript{AHE}$ and $H_{\textrm{c}}$ (Fig.~\ref{fig:fig3}c and Supplementary Fig.~7), H$\textsubscript{Heav}(x)$ is the Heaviside function approximating the switching of $M$, and the Gaussian distribution
\begin{equation}
	G(T^\prime)=\frac{1}{\sqrt{2\pi T_\sigma^2}}\exp \left[-\frac{(T^\prime-T)^2}{2T_\sigma^2}\right],
\label{eq:gaussian}
\end{equation}
characterizes the strain-driven distribution of domains with varying $T_{\textrm{S}}$ by assuming an effective temperature spreading factor $T_\sigma$. As compared in Fig.~\ref{fig:fig5}c, the numerical simulation indeed reproduces qualitatively well the behavior of the minor loops. The observed AHE sign change and the emergence of hump-shpaed Hall features are also present in films grown on SrTiO$_3$(111) (Supplementary Fig.~9). The quality of the substrate/film interface plays a pivotal role in materializing this exquisite tunability of the Berry curvature in Cr$_2$Te$_3$ films. 

In summary, we have discovered several unusual Berry curvature driven effects in the anomalous Hall transport of Cr$_2$Te$_3$ thin films. We report on the growth, detailed magnetic and transport properties of pristine Cr$_2$Te$_3$ MBE thin films deposited on Al$_2$O$_3$(0001) and SrTiO$_3$(111) substrates. A striking sign reversal in the anomalous Hall resistivity, accompanied by a finite magnetization, has been observed and theoretically modeled, revealing the relevance of the nontrivial Berry curvature physics. This unique sign reversal, coupled with the intrinsic strain-modulated magnetic domains in the material, induces a hump-shaped Hall feature in Cr$_2$Te$_3$ thin films. The Berry curvature effect is observed in this case due to the high quality of the substrate/film interface, which is further tunable via different level of strain given by varying film thickness and/or choice of substrates. Our comprehensive experimental and theoretical investigations establish Cr$_2$Te$_3$ to host tunable topological effects related to the intrinsic Berry curvature, thereby providing new perspectives in the field of topological electronics.

\section*{Methods}
\textbf{Sample growth.} The growth of Cr$_2$Te$_3$ thin films, with nominal $t$ ranging from 3 -- 24~u.c.\ were carried out in a MBE system under an ultrahigh-vacuum (UHV) environment of $10^{-10} - 10^{-9}$~Torr. Insulating Al$_2$O$_3$(0001) was primarily used as substrate, whose surface quality was insured by \emph{ex situ} chemical and thermal cleaning and \emph{in situ} outgassing at 800~$^\circ$C for 30~min.\ When using SrTiO$_3$(111), the insulating substrates were first annealed at 930~$^\circ$C for 3~hours in a tube furnace under flowing oxygen environment to achieve passivated surface with atomic flatness and then \emph{in situ} outgassed at 580~$^\circ$C for 30~min. After surface preparation, the substrate temperature was lowered to 230~$^\circ$C for film growth, allowing enough surface mobility for the epitaxial crystallization of the desired phase of Cr$_2$Te$_3$. High-purity (5N) Cr was evaporated from an electron-beam source, while Te was thermally co-evaporated from a Knudsen effusion cell adjusted to maintain a typical Cr:Te flux ratio of 1:10 and a growth rate of approximately 0.005~nm~s$^{-1}$. The epitaxial growth process was monitored by \emph{in situ} RHEED (see Supplementary Fig.~1) operated at 15~kV. The as-grown films were \emph{in situ} annealed at the growth temperature for 30~min, and naturally cooled to room temperature. For \emph{ex situ} characterizations, films were protected by \emph{in situ} capping with Te (2~nm) and AlO$_x$ (10~nm) or Se (20~nm) for later removal for STM measurements. The schematic of film stack is illustrated in Fig.~1f.

\textbf{Structural characterizations.} The XRD patterns were obtained using a parallel beam of Cu K$_{\alpha1}$ radiation with wavelength $\lambda$ = 0.15406~nm in a Rigaku SmartLab system. The $2\theta$ (for OOP measurement) and/or 2$\theta_{\chi}$ (for IP configuration) scan angles were between 10$^\circ$ and 120$^\circ$ with a typical step size of 0.05$^\circ$.\ XRR measurements were performed at the Center for Nanophase Materials Sciences (CNMS), Oak Ridge National Laboratory, on a PANalytical X'Pert Pro MRD equipped with hybrid monochromator and Xe proportional counter. For the XRR measurements, the X-ray beam was generated at 45~kV/40~mA, and the X-ray beam wavelength after the hybrid mirror was $\lambda$ = 0.15406~nm (Cu K$_{\alpha1}$ radiation). To facilitate electron microscopy, planner samples were deposited on Si$_3$N$_4$ TEM grids with thin Sb$_2$Te$_3$ buffer while cross-sectional samples were prepared using focused ion beam (FIB) lift-out method on a Thermo Scientific FEI Quanta 3D dual beam system. STEM imaging was carried out on a Thermo Scientific FEI Titan aberration-corrected system operated at 200~kV. A semi-convergence angle of 17.9~mrad was used. DPC and iDPC images were recorded using a segmented detector. For the 3 u.c. sample, STEM images were acquired with a Themis Z G3 instrument provided by Thermo Fischer Scientific at 200 kV with a beam current of 40 pA and a convergence angle of 25 mrad.

\textbf{Scanning tunneling microscopy.} STM experiments were performed at the Laboratory for Physical Sciences using a home-built low-temperature scanning tunneling microscope \cite{dreyer-2010-rsi} controlled by a Topometrix digital feedback electronic control unit. Samples were loaded into an UHV environment with a base pressure of $5\times10^{-10}$~Torr and heated in front of a residual gas analyzer to verify the removal of the Se capping layer before being transferred to the microscope at 77~K. Scans were performed with an electrochemically-etched tungsten tip and differential spectroscopy data were extracted via a Stanford Research Systems SR830 lock-in amplifier.

\textbf{Polarized neutron reflectometry.}  PNR is a highly penetrating depth-sensitive technique to probe the chemical and magnetic depth profiles with a resolution of 0.5~nm. The depth profiles of the NSLD and MSLD correspond to the depth profile of the chemical and IP magnetization vector distributions on the atomic scale, respectively \cite{pnr-method-1,pnr-method-2,pnr-method-3}. Based on these neutron scattering merits, PNR serves as the powerful technique to simultaneously and nondestructively characterize chemical and magnetic nature of buried interfaces \cite{pnr-method-4}. PNR experiments were performed on the Magnetism Reflectometer at the Spallation Neutron Source at Oak Ridge National Laboratory \cite{pnrValeria, pnrJiang, pnrSyro}, using neutrons with wavelengths $\lambda$ in a band of 0.2 -- 0.8~nm and a high polarization of 98.5--99{\%}. Measurements were conducted in a closed cycle refrigerator (Advanced Research System) equipped with a 1.15~T Bruker electromagnet. Using the time-of-flight method, a collimated polychromatic beam of polarized neutrons with the wavelength band $\delta\lambda$ impinges on the film at a grazing angle $\theta$, interacting with atomic nuclei and the spins of unpaired electrons. The reflected intensity $R^+$ and $R^-$ are measured as a function of momentum transfer, $Q = 4\pi\sin(\theta)/\lambda$, with the neutron spin parallel ($+$) or antiparallel ($-$), respectively, to the applied field. To separate the nuclear from the magnetic scattering, the spin asymmetry ratio SA = $(R^{+}-R^{-})/(R^{+} + R^{-})$ is calculated, for which SA = 0 designating no magnetic moment in the system. Being electrically neutral, spin-polarized neutrons penetrate the entire multilayer structures and probe magnetic and structural composition of the film and buried interfaces down to the substrate.

\textbf{Transport and magnetic measurements.} Electrical transport measurements as a function of of temperature, field and angle were performed in the temperature range of 2 $\--$ 300 K in a Quantum Design Physical Property Measurement System (PPMS) equipped with a 9 T superconducting magnet. A typical ac current ($I_x$) of 5 $\mu$A was injected into the Hall bar ($\sim 0.3\times1.0$ mm$^2$ for hand-scratched or $10\times30$ $\mu$m$^2$ for e-beam patterned) residing in the crystallographic $a$-$b$ plane, while longitudinal ($V_x$) and transverse ($V_y$) voltages were simultaneously monitored using a lock-in technique. For aligning the magnetic field \textbf{H}, a horizontal rotator was used with an angular resolution of $\sim$ 0.1$^\circ$. VSM was used to characterize the magnetization, where linear diamagnetic backgrounds from sample holders/substrates were subtracted to obtain $M(H)$ and $M(T)$. 

\textbf{Theoretical simulations.} First-principles calculations were performed using the Quantum Espresso packages \cite{dft-method-1}.The generalized gradient approximation with the Perdew-Burke-Ernzerhof parameterization (GGA-PBE) was used as the exchange-correlation functional \cite{dft-method-2}. An energy cutoff of 40 Ry and a $6\times6\times4$ $\Gamma$-centered $k$-mesh were applied for the relaxation calculation. The crystal structure of Cr$_2$Te$_3$ was fully optimized until the force on each atom is smaller than 0.05 eV nm$^{-1}$. The optimized lattice constants of bulk Cr$_2$Te$_3$ are $a = b$ = 0.6799 nm and $c$ = 1.2022 nm. For the self-consistent field calculation, SOC was included and a higher $12\times12\times8$ $k$-mesh was used. The magnetization was set along the $-z$ axis. The resulting absolute magnetic moments of the Cr atoms are 3.08, 2.99 and 3.06 $\mu_{\textrm{B}}$ for Cr1, Cr2 and Cr3, respectively. For the Berry curvature and anomalous Hall conductivity calculations, the Wannier90 packages were used \cite{dft-method-3}. Maximally localized Wannier functions (MLWF) including both Cr $d$-orbitals and Te $p$-orbitals were employed to reproduce the DFT-calculated band structure with SOC.

\textbf{Data availability.} The data that support the findings of this study are available from the corresponding author upon reasonable request. 

\section*{Acknowledgments}
The work at MIT was supported by Army Research Office (W911NF-20-2-0061 and DURIP W911NF-20-1-0074), National Science Foundation (NSF-DMR 1700137) and Office of Naval Research (N00014-20-1-2306). H.C. was sponsored by the Army Research Laboratory (ARL) under Cooperative Agreement Number W911NF-19-2-0015. Y.O. and J.S.M. thank the Center for Integrated Quantum Materials (NSF-DMR 1231319) for financial support. D.H. thanks support from NSF grant DMR-1905662 and the Air Force Office of Scientific Research award FA9550-20-1-0247. T.B.E. is partially supported by NSF under Grant No. DGE 1633587. The electron microscopy was performed at the Analytical Instrumentation Facility (AIF) at North Carolina State University, which is supported by the State of North Carolina and NSF (Award No. ECCS-2025064). The AIF is a member of the North Carolina Research Triangle Nanotechnology Network (RTNN), a site in the National Nanotechnology Coordinated Infrastructure (NNCI). This work was carried out with the use of facilities and instrumentation supported by NSF through the Massachusetts Institute of Technology Materials Research Science and Engineering Center DMR - 1419807. This work was carried out in part through the use of MIT.nano's facilities. This project was funded by the MIT-IBM Watson AI Lab. This research used resources at the Spallation Neutron Source, a Department of Energy Office of Science User Facility operated by the Oak Ridge National Laboratory. XRR measurements were conducted at the Center for Nanophase Materials Sciences (CNMS), which is a DOE Office of Science User Facility. STM measurements utilized the facilities and resources of the Laboratory for Physical Sciences. The DFT work was supported in part by the ARL Research Associateship Program (RAP) Cooperative Agreement Number W911NF-16-2-0008 and used the Extreme Science and Engineering Discovery Environment (XSEDE) supported by NSF Grant No. ACI-1548562 and allocation ID TG-DMR130081.

\section*{Author contributions}
H.C., Y.O. and J.S.M. conceived and designed the research. H.C. grew the films with assistance from Y.O., J.S.M., C.R. and P.J.T., carried out magnetization measurements with D.H., collected and analyzed transport data on macroscopic initial Hall bars as well as microdevices fabricated by O.A.V.. T.B.E., W.G., A.C.F. and F.M.R. examined the microstructure using STEM. J.M., M.D. and R.E.B. characterized the surface using STM. J.K. performed XRR measurements, V.L. and H.A. conducted PNR experiments, V.L. analyzed XRR and PNR data. S.K., Y.L., M.R.N., G.Y. and R.K.L. conducted first-principles calculations while A.T.G., G.J.d.C. and P.A.F. simulated micromagnetic responses. H.C., D.H. and J.S.M. wrote the manuscript, with contributions from all authors.

\nocite{*}

\bibliography{0-MS}% Produces the bibliography via BibTeX.

\begin{thebibliography}{53}%
\makeatletter
\providecommand \@ifxundefined [1]{%
 \@ifx{#1\undefined}
}%
\providecommand \@ifnum [1]{%
 \ifnum #1\expandafter \@firstoftwo
 \else \expandafter \@secondoftwo
 \fi
}%
\providecommand \@ifx [1]{%
 \ifx #1\expandafter \@firstoftwo
 \else \expandafter \@secondoftwo
 \fi
}%
\providecommand \natexlab [1]{#1}%
\providecommand \enquote  [1]{``#1''}%
\providecommand \bibnamefont  [1]{#1}%
\providecommand \bibfnamefont [1]{#1}%
\providecommand \citenamefont [1]{#1}%
\providecommand \href@noop [0]{\@secondoftwo}%
\providecommand \href [0]{\begingroup \@sanitize@url \@href}%
\providecommand \@href[1]{\@@startlink{#1}\@@href}%
\providecommand \@@href[1]{\endgroup#1\@@endlink}%
\providecommand \@sanitize@url [0]{\catcode `\\12\catcode `\$12\catcode
  `\&12\catcode `\#12\catcode `\^12\catcode `\_12\catcode `\%12\relax}%
\providecommand \@@startlink[1]{}%
\providecommand \@@endlink[0]{}%
\providecommand \url  [0]{\begingroup\@sanitize@url \@url }%
\providecommand \@url [1]{\endgroup\@href {#1}{\urlprefix }}%
\providecommand \urlprefix  [0]{URL }%
\providecommand \Eprint [0]{\href }%
\providecommand \doibase [0]{https://doi.org/}%
\providecommand \selectlanguage [0]{\@gobble}%
\providecommand \bibinfo  [0]{\@secondoftwo}%
\providecommand \bibfield  [0]{\@secondoftwo}%
\providecommand \translation [1]{[#1]}%
\providecommand \BibitemOpen [0]{}%
\providecommand \bibitemStop [0]{}%
\providecommand \bibitemNoStop [0]{.\EOS\space}%
\providecommand \EOS [0]{\spacefactor3000\relax}%
\providecommand \BibitemShut  [1]{\csname bibitem#1\endcsname}%
\let\auto@bib@innerbib\@empty
%</preamble>
\bibitem [{\citenamefont {Burch}\ \emph {et~al.}(2018)\citenamefont {Burch},
  \citenamefont {Mandrus},\ and\ \citenamefont
  {Park}}]{burch-2018-review-nature}%
  \BibitemOpen
  \bibfield  {author} {\bibinfo {author} {\bibfnamefont {K.~S.}\ \bibnamefont
  {Burch}}, \bibinfo {author} {\bibfnamefont {D.}~\bibnamefont {Mandrus}},\
  and\ \bibinfo {author} {\bibfnamefont {J.-G.}\ \bibnamefont {Park}},\
  }\bibfield  {title} {\bibinfo {title} {{Magnetism in two-dimensional van der
  Waals materials}},\ }\href {https://doi.org/10.1038/s41586-018-0631-z}
  {\bibfield  {journal} {\bibinfo  {journal} {Nature}\ }\textbf {\bibinfo
  {volume} {563}},\ \bibinfo {pages} {47} (\bibinfo {year} {2018})}\BibitemShut
  {NoStop}%
\bibitem [{\citenamefont {Ipser}\ \emph {et~al.}(1983)\citenamefont {Ipser},
  \citenamefont {Komarek},\ and\ \citenamefont {Klepp}}]{Ipser-1983-JLCM}%
  \BibitemOpen
  \bibfield  {author} {\bibinfo {author} {\bibfnamefont {H.}~\bibnamefont
  {Ipser}}, \bibinfo {author} {\bibfnamefont {K.~L.}\ \bibnamefont {Komarek}},\
  and\ \bibinfo {author} {\bibfnamefont {K.~O.}\ \bibnamefont {Klepp}},\
  }\bibfield  {title} {\bibinfo {title} {{Transition metal-chalcogen systems
  viii: The Cr-Te phase diagram}},\ }\href
  {https://doi.org/10.1016/0022-5088(83)90493-9} {\bibfield  {journal}
  {\bibinfo  {journal} {Journal of the Less Common Metals}\ }\textbf {\bibinfo
  {volume} {92}},\ \bibinfo {pages} {265} (\bibinfo {year} {1983})}\BibitemShut
  {NoStop}%
\bibitem [{\citenamefont {McGuire}\ \emph {et~al.}(2017)\citenamefont
  {McGuire}, \citenamefont {Garlea}, \citenamefont {Kc}, \citenamefont
  {Cooper}, \citenamefont {Yan}, \citenamefont {Cao},\ and\ \citenamefont
  {Sales}}]{McGuire-2017-PRB-0.333}%
  \BibitemOpen
  \bibfield  {author} {\bibinfo {author} {\bibfnamefont {M.~A.}\ \bibnamefont
  {McGuire}}, \bibinfo {author} {\bibfnamefont {V.~O.}\ \bibnamefont {Garlea}},
  \bibinfo {author} {\bibfnamefont {S.}~\bibnamefont {Kc}}, \bibinfo {author}
  {\bibfnamefont {V.~R.}\ \bibnamefont {Cooper}}, \bibinfo {author}
  {\bibfnamefont {J.}~\bibnamefont {Yan}}, \bibinfo {author} {\bibfnamefont
  {H.}~\bibnamefont {Cao}},\ and\ \bibinfo {author} {\bibfnamefont {B.~C.}\
  \bibnamefont {Sales}},\ }\bibfield  {title} {\bibinfo {title}
  {{Antiferromagnetism in the van der Waals layered spin-lozenge semiconductor
  CrTe$_3$}},\ }\href {https://doi.org/10.1103/PhysRevB.95.144421} {\bibfield
  {journal} {\bibinfo  {journal} {Physical Review B}\ }\textbf {\bibinfo
  {volume} {95}},\ \bibinfo {pages} {144421} (\bibinfo {year}
  {2017})}\BibitemShut {NoStop}%
\bibitem [{\citenamefont {Zhang}\ \emph
  {et~al.}(2021{\natexlab{a}})\citenamefont {Zhang}, \citenamefont {Lu},
  \citenamefont {Liu}, \citenamefont {Niu}, \citenamefont {Sun}, \citenamefont
  {Cook}, \citenamefont {Vaninger}, \citenamefont {Miceli}, \citenamefont
  {Singh}, \citenamefont {Lian}, \citenamefont {Chang}, \citenamefont {He},
  \citenamefont {Du}, \citenamefont {He}, \citenamefont {Zhang}, \citenamefont
  {Bian},\ and\ \citenamefont {Xu}}]{Zhang-2021-NC-0.5}%
  \BibitemOpen
  \bibfield  {author} {\bibinfo {author} {\bibfnamefont {X.}~\bibnamefont
  {Zhang}}, \bibinfo {author} {\bibfnamefont {Q.}~\bibnamefont {Lu}}, \bibinfo
  {author} {\bibfnamefont {W.}~\bibnamefont {Liu}}, \bibinfo {author}
  {\bibfnamefont {W.}~\bibnamefont {Niu}}, \bibinfo {author} {\bibfnamefont
  {J.}~\bibnamefont {Sun}}, \bibinfo {author} {\bibfnamefont {J.}~\bibnamefont
  {Cook}}, \bibinfo {author} {\bibfnamefont {M.}~\bibnamefont {Vaninger}},
  \bibinfo {author} {\bibfnamefont {P.~F.}\ \bibnamefont {Miceli}}, \bibinfo
  {author} {\bibfnamefont {D.~J.}\ \bibnamefont {Singh}}, \bibinfo {author}
  {\bibfnamefont {S.-W.}\ \bibnamefont {Lian}}, \bibinfo {author}
  {\bibfnamefont {T.-R.}\ \bibnamefont {Chang}}, \bibinfo {author}
  {\bibfnamefont {X.}~\bibnamefont {He}}, \bibinfo {author} {\bibfnamefont
  {J.}~\bibnamefont {Du}}, \bibinfo {author} {\bibfnamefont {L.}~\bibnamefont
  {He}}, \bibinfo {author} {\bibfnamefont {R.}~\bibnamefont {Zhang}}, \bibinfo
  {author} {\bibfnamefont {G.}~\bibnamefont {Bian}},\ and\ \bibinfo {author}
  {\bibfnamefont {Y.}~\bibnamefont {Xu}},\ }\bibfield  {title} {\bibinfo
  {title} {{Room-temperature intrinsic ferromagnetism in epitaxial CrTe$_2$
  ultrathin films}},\ }\href {https://doi.org/10.1038/s41467-021-22777-x}
  {\bibfield  {journal} {\bibinfo  {journal} {Nature Communications}\ }\textbf
  {\bibinfo {volume} {12}},\ \bibinfo {pages} {2492} (\bibinfo {year}
  {2021}{\natexlab{a}})}\BibitemShut {NoStop}%
\bibitem [{\citenamefont {Liu}\ and\ \citenamefont
  {Petrovic}(2018)}]{Liu-2018-PRB-0.625}%
  \BibitemOpen
  \bibfield  {author} {\bibinfo {author} {\bibfnamefont {Y.}~\bibnamefont
  {Liu}}\ and\ \bibinfo {author} {\bibfnamefont {C.}~\bibnamefont {Petrovic}},\
  }\bibfield  {title} {\bibinfo {title} {{Anomalous Hall effect in the trigonal
  Cr$_5$Te$_8$ single crystal}},\ }\href
  {https://doi.org/10.1103/PhysRevB.98.195122} {\bibfield  {journal} {\bibinfo
  {journal} {Physical Review B}\ }\textbf {\bibinfo {volume} {98}},\ \bibinfo
  {pages} {195122} (\bibinfo {year} {2018})}\BibitemShut {NoStop}%
\bibitem [{\citenamefont {Wen}\ \emph {et~al.}(2020)\citenamefont {Wen},
  \citenamefont {Liu}, \citenamefont {Zhang}, \citenamefont {Xia},
  \citenamefont {Zhai}, \citenamefont {Zhang}, \citenamefont {Zhai},
  \citenamefont {Shen}, \citenamefont {He}, \citenamefont {Cheng},
  \citenamefont {Yin}, \citenamefont {Yao}, \citenamefont {Getaye~Sendeku},
  \citenamefont {Wang}, \citenamefont {Ye}, \citenamefont {Liu}, \citenamefont
  {Jiang}, \citenamefont {Shan}, \citenamefont {Long},\ and\ \citenamefont
  {He}}]{Wen-2020-NL-0.667}%
  \BibitemOpen
  \bibfield  {author} {\bibinfo {author} {\bibfnamefont {Y.}~\bibnamefont
  {Wen}}, \bibinfo {author} {\bibfnamefont {Z.}~\bibnamefont {Liu}}, \bibinfo
  {author} {\bibfnamefont {Y.}~\bibnamefont {Zhang}}, \bibinfo {author}
  {\bibfnamefont {C.}~\bibnamefont {Xia}}, \bibinfo {author} {\bibfnamefont
  {B.}~\bibnamefont {Zhai}}, \bibinfo {author} {\bibfnamefont {X.}~\bibnamefont
  {Zhang}}, \bibinfo {author} {\bibfnamefont {G.}~\bibnamefont {Zhai}},
  \bibinfo {author} {\bibfnamefont {C.}~\bibnamefont {Shen}}, \bibinfo {author}
  {\bibfnamefont {P.}~\bibnamefont {He}}, \bibinfo {author} {\bibfnamefont
  {R.}~\bibnamefont {Cheng}}, \bibinfo {author} {\bibfnamefont
  {L.}~\bibnamefont {Yin}}, \bibinfo {author} {\bibfnamefont {Y.}~\bibnamefont
  {Yao}}, \bibinfo {author} {\bibfnamefont {M.}~\bibnamefont {Getaye~Sendeku}},
  \bibinfo {author} {\bibfnamefont {Z.}~\bibnamefont {Wang}}, \bibinfo {author}
  {\bibfnamefont {X.}~\bibnamefont {Ye}}, \bibinfo {author} {\bibfnamefont
  {C.}~\bibnamefont {Liu}}, \bibinfo {author} {\bibfnamefont {C.}~\bibnamefont
  {Jiang}}, \bibinfo {author} {\bibfnamefont {C.}~\bibnamefont {Shan}},
  \bibinfo {author} {\bibfnamefont {Y.}~\bibnamefont {Long}},\ and\ \bibinfo
  {author} {\bibfnamefont {J.}~\bibnamefont {He}},\ }\bibfield  {title}
  {\bibinfo {title} {{Tunable Room-Temperature Ferromagnetism in
  Two-Dimensional Cr$_2$Te$_3$}},\ }\href
  {https://doi.org/10.1021/acs.nanolett.9b05128} {\bibfield  {journal}
  {\bibinfo  {journal} {Nano Letters}\ }\textbf {\bibinfo {volume} {20}},\
  \bibinfo {pages} {3130} (\bibinfo {year} {2020})}\BibitemShut {NoStop}%
\bibitem [{\citenamefont {Cao}\ \emph {et~al.}(2019)\citenamefont {Cao},
  \citenamefont {Zhang}, \citenamefont {Frontzek}, \citenamefont {Xie},
  \citenamefont {Gong}, \citenamefont {Sterbinsky},\ and\ \citenamefont
  {Jin}}]{Cao-2019-PRM-0.75x}%
  \BibitemOpen
  \bibfield  {author} {\bibinfo {author} {\bibfnamefont {G.}~\bibnamefont
  {Cao}}, \bibinfo {author} {\bibfnamefont {Q.}~\bibnamefont {Zhang}}, \bibinfo
  {author} {\bibfnamefont {M.}~\bibnamefont {Frontzek}}, \bibinfo {author}
  {\bibfnamefont {W.}~\bibnamefont {Xie}}, \bibinfo {author} {\bibfnamefont
  {D.}~\bibnamefont {Gong}}, \bibinfo {author} {\bibfnamefont {G.~E.}\
  \bibnamefont {Sterbinsky}},\ and\ \bibinfo {author} {\bibfnamefont
  {R.}~\bibnamefont {Jin}},\ }\bibfield  {title} {\bibinfo {title} {{Structure,
  chromium vacancies, and magnetism in a Cr$_{12-x}$Te$_{16}$ compound}},\
  }\href {https://doi.org/10.1103/PhysRevMaterials.3.125001} {\bibfield
  {journal} {\bibinfo  {journal} {Physical Review Materials}\ }\textbf
  {\bibinfo {volume} {3}},\ \bibinfo {pages} {125001} (\bibinfo {year}
  {2019})}\BibitemShut {NoStop}%
\bibitem [{\citenamefont {Chua}\ \emph {et~al.}(2021)\citenamefont {Chua},
  \citenamefont {Zhou}, \citenamefont {Yu}, \citenamefont {Yu}, \citenamefont
  {Gou}, \citenamefont {Zhu}, \citenamefont {Zhang}, \citenamefont {Liu},
  \citenamefont {Breese}, \citenamefont {Chen}, \citenamefont {Loh},
  \citenamefont {Feng}, \citenamefont {Yang}, \citenamefont {Huang},\ and\
  \citenamefont {Wee}}]{Chua-2021-AM-0.75}%
  \BibitemOpen
  \bibfield  {author} {\bibinfo {author} {\bibfnamefont {R.}~\bibnamefont
  {Chua}}, \bibinfo {author} {\bibfnamefont {J.}~\bibnamefont {Zhou}}, \bibinfo
  {author} {\bibfnamefont {X.}~\bibnamefont {Yu}}, \bibinfo {author}
  {\bibfnamefont {W.}~\bibnamefont {Yu}}, \bibinfo {author} {\bibfnamefont
  {J.}~\bibnamefont {Gou}}, \bibinfo {author} {\bibfnamefont {R.}~\bibnamefont
  {Zhu}}, \bibinfo {author} {\bibfnamefont {L.}~\bibnamefont {Zhang}}, \bibinfo
  {author} {\bibfnamefont {M.}~\bibnamefont {Liu}}, \bibinfo {author}
  {\bibfnamefont {M.~B.~H.}\ \bibnamefont {Breese}}, \bibinfo {author}
  {\bibfnamefont {W.}~\bibnamefont {Chen}}, \bibinfo {author} {\bibfnamefont
  {K.~P.}\ \bibnamefont {Loh}}, \bibinfo {author} {\bibfnamefont {Y.~P.}\
  \bibnamefont {Feng}}, \bibinfo {author} {\bibfnamefont {M.}~\bibnamefont
  {Yang}}, \bibinfo {author} {\bibfnamefont {Y.~L.}\ \bibnamefont {Huang}},\
  and\ \bibinfo {author} {\bibfnamefont {A.~T.~S.}\ \bibnamefont {Wee}},\
  }\bibfield  {title} {\bibinfo {title} {{Room Temperature Ferromagnetism of
  Monolayer Chromium Telluride with Perpendicular Magnetic Anisotropy}},\
  }\href {https://doi.org/10.1002/adma.202103360} {\bibfield  {journal}
  {\bibinfo  {journal} {Advanced Materials}\ }\textbf {\bibinfo {volume}
  {33}},\ \bibinfo {pages} {2103360} (\bibinfo {year} {2021})}\BibitemShut
  {NoStop}%
\bibitem [{\citenamefont {Gong}\ and\ \citenamefont {Zhang}(2019)}]{bib3}%
  \BibitemOpen
  \bibfield  {author} {\bibinfo {author} {\bibfnamefont {C.}~\bibnamefont
  {Gong}}\ and\ \bibinfo {author} {\bibfnamefont {X.}~\bibnamefont {Zhang}},\
  }\bibfield  {title} {\bibinfo {title} {{Two-dimensional magnetic crystals and
  emergent heterostructure devices}},\ }\href
  {https://doi.org/10.1126/science.aav4450} {\bibfield  {journal} {\bibinfo
  {journal} {Science}\ }\textbf {\bibinfo {volume} {363}},\ \bibinfo {pages}
  {eaav4450} (\bibinfo {year} {2019})}\BibitemShut {NoStop}%
\bibitem [{\citenamefont {Hasan}\ and\ \citenamefont {Kane}(2010)}]{bib4}%
  \BibitemOpen
  \bibfield  {author} {\bibinfo {author} {\bibfnamefont {M.~Z.}\ \bibnamefont
  {Hasan}}\ and\ \bibinfo {author} {\bibfnamefont {C.~L.}\ \bibnamefont
  {Kane}},\ }\bibfield  {title} {\bibinfo {title} {{Colloquium: Topological
  insulators}},\ }\href {https://doi.org/10.1103/RevModPhys.82.3045} {\bibfield
   {journal} {\bibinfo  {journal} {Reviews of Modern Physics}\ }\textbf
  {\bibinfo {volume} {82}},\ \bibinfo {pages} {3045} (\bibinfo {year}
  {2010})}\BibitemShut {NoStop}%
\bibitem [{\citenamefont {Armitage}\ \emph {et~al.}(2018)\citenamefont
  {Armitage}, \citenamefont {Mele},\ and\ \citenamefont
  {Vishwanath}}]{armitage-2018-rmp}%
  \BibitemOpen
  \bibfield  {author} {\bibinfo {author} {\bibfnamefont {N.~P.}\ \bibnamefont
  {Armitage}}, \bibinfo {author} {\bibfnamefont {E.~J.}\ \bibnamefont {Mele}},\
  and\ \bibinfo {author} {\bibfnamefont {A.}~\bibnamefont {Vishwanath}},\
  }\bibfield  {title} {\bibinfo {title} {{Weyl and Dirac semimetals in
  three-dimensional solids}},\ }\href
  {https://doi.org/10.1103/RevModPhys.90.015001} {\bibfield  {journal}
  {\bibinfo  {journal} {Reviews of Modern Physics}\ }\textbf {\bibinfo {volume}
  {90}},\ \bibinfo {pages} {15001} (\bibinfo {year} {2018})}\BibitemShut
  {NoStop}%
\bibitem [{\citenamefont {Jiang}\ \emph {et~al.}(2020)\citenamefont {Jiang},
  \citenamefont {Xiao}, \citenamefont {Wang}, \citenamefont {Shin},
  \citenamefont {Andreoli}, \citenamefont {Zhang}, \citenamefont {Xiao},
  \citenamefont {Zhao}, \citenamefont {Kayyalha}, \citenamefont {Zhang},
  \citenamefont {Wang}, \citenamefont {Zang}, \citenamefont {Liu},
  \citenamefont {Samarth}, \citenamefont {Chan},\ and\ \citenamefont
  {Chang}}]{Jiang-2020-NM}%
  \BibitemOpen
  \bibfield  {author} {\bibinfo {author} {\bibfnamefont {J.}~\bibnamefont
  {Jiang}}, \bibinfo {author} {\bibfnamefont {D.}~\bibnamefont {Xiao}},
  \bibinfo {author} {\bibfnamefont {F.}~\bibnamefont {Wang}}, \bibinfo {author}
  {\bibfnamefont {J.-H.}\ \bibnamefont {Shin}}, \bibinfo {author}
  {\bibfnamefont {D.}~\bibnamefont {Andreoli}}, \bibinfo {author}
  {\bibfnamefont {J.}~\bibnamefont {Zhang}}, \bibinfo {author} {\bibfnamefont
  {R.}~\bibnamefont {Xiao}}, \bibinfo {author} {\bibfnamefont {Y.-F.}\
  \bibnamefont {Zhao}}, \bibinfo {author} {\bibfnamefont {M.}~\bibnamefont
  {Kayyalha}}, \bibinfo {author} {\bibfnamefont {L.}~\bibnamefont {Zhang}},
  \bibinfo {author} {\bibfnamefont {K.}~\bibnamefont {Wang}}, \bibinfo {author}
  {\bibfnamefont {J.}~\bibnamefont {Zang}}, \bibinfo {author} {\bibfnamefont
  {C.}~\bibnamefont {Liu}}, \bibinfo {author} {\bibfnamefont {N.}~\bibnamefont
  {Samarth}}, \bibinfo {author} {\bibfnamefont {M.~H.~W.}\ \bibnamefont
  {Chan}},\ and\ \bibinfo {author} {\bibfnamefont {C.-Z.}\ \bibnamefont
  {Chang}},\ }\bibfield  {title} {\bibinfo {title} {{Concurrence of quantum
  anomalous Hall and topological Hall effects in magnetic topological insulator
  sandwich heterostructures}},\ }\href
  {https://doi.org/10.1038/s41563-020-0605-z} {\bibfield  {journal} {\bibinfo
  {journal} {Nature Materials}\ }\textbf {\bibinfo {volume} {19}},\ \bibinfo
  {pages} {732} (\bibinfo {year} {2020})}\BibitemShut {NoStop}%
\bibitem [{\citenamefont {Bernevig}\ \emph {et~al.}(2022)\citenamefont
  {Bernevig}, \citenamefont {Felser},\ and\ \citenamefont
  {Beidenkopf}}]{Bernevig-2022-review-nature}%
  \BibitemOpen
  \bibfield  {author} {\bibinfo {author} {\bibfnamefont {B.~A.}\ \bibnamefont
  {Bernevig}}, \bibinfo {author} {\bibfnamefont {C.}~\bibnamefont {Felser}},\
  and\ \bibinfo {author} {\bibfnamefont {H.}~\bibnamefont {Beidenkopf}},\
  }\bibfield  {title} {\bibinfo {title} {{Progress and prospects in magnetic
  topological materials}},\ }\href {https://doi.org/10.1038/s41586-021-04105-x}
  {\bibfield  {journal} {\bibinfo  {journal} {Nature}\ }\textbf {\bibinfo
  {volume} {603}},\ \bibinfo {pages} {41} (\bibinfo {year} {2022})}\BibitemShut
  {NoStop}%
\bibitem [{\citenamefont {Chi}\ and\ \citenamefont {Moodera}(2022)}]{Chi-QAH}%
  \BibitemOpen
  \bibfield  {author} {\bibinfo {author} {\bibfnamefont {H.}~\bibnamefont
  {Chi}}\ and\ \bibinfo {author} {\bibfnamefont {J.~S.}\ \bibnamefont
  {Moodera}},\ }\bibfield  {title} {\bibinfo {title} {Progress and prospects in
  the quantum anomalous hall effect},\ }\href
  {https://doi.org/10.1063/5.0100989} {\bibfield  {journal} {\bibinfo
  {journal} {APL Materials}\ }\textbf {\bibinfo {volume} {10}},\ \bibinfo
  {pages} {090903} (\bibinfo {year} {2022})}\BibitemShut {NoStop}%
\bibitem [{\citenamefont {{\v{Z}}uti{\'{c}}}\ \emph {et~al.}(2004)\citenamefont
  {{\v{Z}}uti{\'{c}}}, \citenamefont {Fabian},\ and\ \citenamefont
  {Das~Sarma}}]{bib6}%
  \BibitemOpen
  \bibfield  {author} {\bibinfo {author} {\bibfnamefont {I.}~\bibnamefont
  {{\v{Z}}uti{\'{c}}}}, \bibinfo {author} {\bibfnamefont {J.}~\bibnamefont
  {Fabian}},\ and\ \bibinfo {author} {\bibfnamefont {S.}~\bibnamefont
  {Das~Sarma}},\ }\bibfield  {title} {\bibinfo {title} {{Spintronics:
  Fundamentals and applications}},\ }\href
  {https://doi.org/10.1103/RevModPhys.76.323} {\bibfield  {journal} {\bibinfo
  {journal} {Reviews of Modern Physics}\ }\textbf {\bibinfo {volume} {76}},\
  \bibinfo {pages} {323} (\bibinfo {year} {2004})}\BibitemShut {NoStop}%
\bibitem [{\citenamefont {Tokura}\ \emph {et~al.}(2019)\citenamefont {Tokura},
  \citenamefont {Yasuda},\ and\ \citenamefont {Tsukazaki}}]{bib7}%
  \BibitemOpen
  \bibfield  {author} {\bibinfo {author} {\bibfnamefont {Y.}~\bibnamefont
  {Tokura}}, \bibinfo {author} {\bibfnamefont {K.}~\bibnamefont {Yasuda}},\
  and\ \bibinfo {author} {\bibfnamefont {A.}~\bibnamefont {Tsukazaki}},\
  }\bibfield  {title} {\bibinfo {title} {{Magnetic topological insulators}},\
  }\href {https://doi.org/10.1038/s42254-018-0011-5} {\bibfield  {journal}
  {\bibinfo  {journal} {Nature Reviews Physics}\ }\textbf {\bibinfo {volume}
  {1}},\ \bibinfo {pages} {126} (\bibinfo {year} {2019})}\BibitemShut {NoStop}%
\bibitem [{\citenamefont {Karplus}\ and\ \citenamefont
  {Luttinger}(1954)}]{bib8}%
  \BibitemOpen
  \bibfield  {author} {\bibinfo {author} {\bibfnamefont {R.}~\bibnamefont
  {Karplus}}\ and\ \bibinfo {author} {\bibfnamefont {J.~M.}\ \bibnamefont
  {Luttinger}},\ }\bibfield  {title} {\bibinfo {title} {{Hall Effect in
  Ferromagnetics}},\ }\href {https://doi.org/10.1103/PhysRev.95.1154}
  {\bibfield  {journal} {\bibinfo  {journal} {Physical Review}\ }\textbf
  {\bibinfo {volume} {95}},\ \bibinfo {pages} {1154} (\bibinfo {year}
  {1954})}\BibitemShut {NoStop}%
\bibitem [{\citenamefont {Berry}(1984)}]{bib9}%
  \BibitemOpen
  \bibfield  {author} {\bibinfo {author} {\bibfnamefont {M.~V.}\ \bibnamefont
  {Berry}},\ }\bibfield  {title} {\bibinfo {title} {{Quantal phase factors
  accompanying adiabatic changes}},\ }\href
  {https://doi.org/10.1098/rspa.1984.0023} {\bibfield  {journal} {\bibinfo
  {journal} {Proceedings of the Royal Society of London. A. Mathematical and
  Physical Sciences}\ }\textbf {\bibinfo {volume} {392}},\ \bibinfo {pages}
  {45} (\bibinfo {year} {1984})}\BibitemShut {NoStop}%
\bibitem [{\citenamefont {Nagaosa}\ \emph {et~al.}(2010)\citenamefont
  {Nagaosa}, \citenamefont {Sinova}, \citenamefont {Onoda}, \citenamefont
  {MacDonald},\ and\ \citenamefont {Ong}}]{bib10}%
  \BibitemOpen
  \bibfield  {author} {\bibinfo {author} {\bibfnamefont {N.}~\bibnamefont
  {Nagaosa}}, \bibinfo {author} {\bibfnamefont {J.}~\bibnamefont {Sinova}},
  \bibinfo {author} {\bibfnamefont {S.}~\bibnamefont {Onoda}}, \bibinfo
  {author} {\bibfnamefont {A.~H.}\ \bibnamefont {MacDonald}},\ and\ \bibinfo
  {author} {\bibfnamefont {N.~P.}\ \bibnamefont {Ong}},\ }\bibfield  {title}
  {\bibinfo {title} {{Anomalous Hall effect}},\ }\href
  {https://doi.org/10.1103/RevModPhys.82.1539} {\bibfield  {journal} {\bibinfo
  {journal} {Reviews of Modern Physics}\ }\textbf {\bibinfo {volume} {82}},\
  \bibinfo {pages} {1539} (\bibinfo {year} {2010})}\BibitemShut {NoStop}%
\bibitem [{\citenamefont {Machida}\ \emph {et~al.}(2010)\citenamefont
  {Machida}, \citenamefont {Nakatsuji}, \citenamefont {Onoda}, \citenamefont
  {Tayama},\ and\ \citenamefont {Sakakibara}}]{bib11}%
  \BibitemOpen
  \bibfield  {author} {\bibinfo {author} {\bibfnamefont {Y.}~\bibnamefont
  {Machida}}, \bibinfo {author} {\bibfnamefont {S.}~\bibnamefont {Nakatsuji}},
  \bibinfo {author} {\bibfnamefont {S.}~\bibnamefont {Onoda}}, \bibinfo
  {author} {\bibfnamefont {T.}~\bibnamefont {Tayama}},\ and\ \bibinfo {author}
  {\bibfnamefont {T.}~\bibnamefont {Sakakibara}},\ }\bibfield  {title}
  {\bibinfo {title} {{Time-reversal symmetry breaking and spontaneous Hall
  effect without magnetic dipole order}},\ }\href
  {https://doi.org/10.1038/nature08680} {\bibfield  {journal} {\bibinfo
  {journal} {Nature}\ }\textbf {\bibinfo {volume} {463}},\ \bibinfo {pages}
  {210} (\bibinfo {year} {2010})}\BibitemShut {NoStop}%
\bibitem [{\citenamefont {{\v{S}mejkal, Libor and MacDonald, Allan H. and
  Sinova, Jairo and Nakatsuji, Satoru and Jungwirth, Tomas}}(2022)}]{bib12}%
  \BibitemOpen
  \bibfield  {author} {\bibinfo {author} {\bibnamefont {{\v{S}mejkal, Libor and
  MacDonald, Allan H. and Sinova, Jairo and Nakatsuji, Satoru and Jungwirth,
  Tomas}}},\ }\bibfield  {title} {\bibinfo {title} {{Anomalous Hall
  antiferromagnets}},\ }\href {https://doi.org/10.1038/s41578-022-00430-3}
  {\bibfield  {journal} {\bibinfo  {journal} {Nature Reviews Materials}\
  }\textbf {\bibinfo {volume} {7}},\ \bibinfo {pages} {482} (\bibinfo {year}
  {2022})}\BibitemShut {NoStop}%
\bibitem [{\citenamefont {Liang}\ \emph {et~al.}(2018)\citenamefont {Liang},
  \citenamefont {Lin}, \citenamefont {Gibson}, \citenamefont {Kushwaha},
  \citenamefont {Liu}, \citenamefont {Wang}, \citenamefont {Xiong},
  \citenamefont {Sobota}, \citenamefont {Hashimoto}, \citenamefont {Kirchmann},
  \citenamefont {Shen}, \citenamefont {Cava},\ and\ \citenamefont
  {Ong}}]{bib13}%
  \BibitemOpen
  \bibfield  {author} {\bibinfo {author} {\bibfnamefont {T.}~\bibnamefont
  {Liang}}, \bibinfo {author} {\bibfnamefont {J.}~\bibnamefont {Lin}}, \bibinfo
  {author} {\bibfnamefont {Q.}~\bibnamefont {Gibson}}, \bibinfo {author}
  {\bibfnamefont {S.}~\bibnamefont {Kushwaha}}, \bibinfo {author}
  {\bibfnamefont {M.}~\bibnamefont {Liu}}, \bibinfo {author} {\bibfnamefont
  {W.}~\bibnamefont {Wang}}, \bibinfo {author} {\bibfnamefont {H.}~\bibnamefont
  {Xiong}}, \bibinfo {author} {\bibfnamefont {J.~A.}\ \bibnamefont {Sobota}},
  \bibinfo {author} {\bibfnamefont {M.}~\bibnamefont {Hashimoto}}, \bibinfo
  {author} {\bibfnamefont {P.~S.}\ \bibnamefont {Kirchmann}}, \bibinfo {author}
  {\bibfnamefont {Z.-X.}\ \bibnamefont {Shen}}, \bibinfo {author}
  {\bibfnamefont {R.~J.}\ \bibnamefont {Cava}},\ and\ \bibinfo {author}
  {\bibfnamefont {N.~P.}\ \bibnamefont {Ong}},\ }\bibfield  {title} {\bibinfo
  {title} {{Anomalous Hall effect in ZrTe$_5$}},\ }\href
  {https://doi.org/10.1038/s41567-018-0078-z} {\bibfield  {journal} {\bibinfo
  {journal} {Nature Physics}\ }\textbf {\bibinfo {volume} {14}},\ \bibinfo
  {pages} {451} (\bibinfo {year} {2018})}\BibitemShut {NoStop}%
\bibitem [{\citenamefont {Fang}\ \emph {et~al.}(2003)\citenamefont {Fang},
  \citenamefont {Nagaosa}, \citenamefont {Takahashi}, \citenamefont {Asamitsu},
  \citenamefont {Mathieu}, \citenamefont {Ogasawara}, \citenamefont {Yamada},
  \citenamefont {Kawasaki}, \citenamefont {Tokura},\ and\ \citenamefont
  {Terakura}}]{bib14}%
  \BibitemOpen
  \bibfield  {author} {\bibinfo {author} {\bibfnamefont {Z.}~\bibnamefont
  {Fang}}, \bibinfo {author} {\bibfnamefont {N.}~\bibnamefont {Nagaosa}},
  \bibinfo {author} {\bibfnamefont {K.~S.}\ \bibnamefont {Takahashi}}, \bibinfo
  {author} {\bibfnamefont {A.}~\bibnamefont {Asamitsu}}, \bibinfo {author}
  {\bibfnamefont {R.}~\bibnamefont {Mathieu}}, \bibinfo {author} {\bibfnamefont
  {T.}~\bibnamefont {Ogasawara}}, \bibinfo {author} {\bibfnamefont
  {H.}~\bibnamefont {Yamada}}, \bibinfo {author} {\bibfnamefont
  {M.}~\bibnamefont {Kawasaki}}, \bibinfo {author} {\bibfnamefont
  {Y.}~\bibnamefont {Tokura}},\ and\ \bibinfo {author} {\bibfnamefont
  {K.}~\bibnamefont {Terakura}},\ }\bibfield  {title} {\bibinfo {title} {{The
  Anomalous Hall Effect and Magnetic Monopoles in Momentum Space}},\ }\href
  {https://doi.org/10.1126/science.1089408} {\bibfield  {journal} {\bibinfo
  {journal} {Science}\ }\textbf {\bibinfo {volume} {302}},\ \bibinfo {pages}
  {92} (\bibinfo {year} {2003})}\BibitemShut {NoStop}%
\bibitem [{\citenamefont {Kim}\ \emph {et~al.}(2018)\citenamefont {Kim},
  \citenamefont {Seo}, \citenamefont {Lee}, \citenamefont {Ko}, \citenamefont
  {Kim}, \citenamefont {Jang}, \citenamefont {Ok}, \citenamefont {Lee},
  \citenamefont {Jo}, \citenamefont {Kang}, \citenamefont {Shim}, \citenamefont
  {Kim}, \citenamefont {Yeom}, \citenamefont {Il~Min}, \citenamefont {Yang},\
  and\ \citenamefont {Kim}}]{bib15}%
  \BibitemOpen
  \bibfield  {author} {\bibinfo {author} {\bibfnamefont {K.}~\bibnamefont
  {Kim}}, \bibinfo {author} {\bibfnamefont {J.}~\bibnamefont {Seo}}, \bibinfo
  {author} {\bibfnamefont {E.}~\bibnamefont {Lee}}, \bibinfo {author}
  {\bibfnamefont {K.~T.}\ \bibnamefont {Ko}}, \bibinfo {author} {\bibfnamefont
  {B.~S.}\ \bibnamefont {Kim}}, \bibinfo {author} {\bibfnamefont {B.~G.}\
  \bibnamefont {Jang}}, \bibinfo {author} {\bibfnamefont {J.~M.}\ \bibnamefont
  {Ok}}, \bibinfo {author} {\bibfnamefont {J.}~\bibnamefont {Lee}}, \bibinfo
  {author} {\bibfnamefont {Y.~J.}\ \bibnamefont {Jo}}, \bibinfo {author}
  {\bibfnamefont {W.}~\bibnamefont {Kang}}, \bibinfo {author} {\bibfnamefont
  {J.~H.}\ \bibnamefont {Shim}}, \bibinfo {author} {\bibfnamefont
  {C.}~\bibnamefont {Kim}}, \bibinfo {author} {\bibfnamefont {H.~W.}\
  \bibnamefont {Yeom}}, \bibinfo {author} {\bibfnamefont {B.}~\bibnamefont
  {Il~Min}}, \bibinfo {author} {\bibfnamefont {B.-J.}\ \bibnamefont {Yang}},\
  and\ \bibinfo {author} {\bibfnamefont {J.~S.}\ \bibnamefont {Kim}},\
  }\bibfield  {title} {\bibinfo {title} {{Large anomalous Hall current induced
  by topological nodal lines in a ferromagnetic van der Waals semimetal}},\
  }\href {https://doi.org/10.1038/s41563-018-0132-3} {\bibfield  {journal}
  {\bibinfo  {journal} {Nature Materials}\ }\textbf {\bibinfo {volume} {17}},\
  \bibinfo {pages} {794} (\bibinfo {year} {2018})}\BibitemShut {NoStop}%
\bibitem [{\citenamefont {Yuzuri}\ \emph {et~al.}(1987)\citenamefont {Yuzuri},
  \citenamefont {Kanomata},\ and\ \citenamefont {Kaneko}}]{bib16}%
  \BibitemOpen
  \bibfield  {author} {\bibinfo {author} {\bibfnamefont {M.}~\bibnamefont
  {Yuzuri}}, \bibinfo {author} {\bibfnamefont {T.}~\bibnamefont {Kanomata}},\
  and\ \bibinfo {author} {\bibfnamefont {T.}~\bibnamefont {Kaneko}},\
  }\bibfield  {title} {\bibinfo {title} {{The pressure effect on the Curie
  temperature and exchange striction of Cr$_2$S$_3$ and Cr$_2$Te$_3$}},\ }\href
  {https://doi.org/10.1016/0304-8853(87)90416-1} {\bibfield  {journal}
  {\bibinfo  {journal} {Journal of Magnetism and Magnetic Materials}\ }\textbf
  {\bibinfo {volume} {70}},\ \bibinfo {pages} {223} (\bibinfo {year}
  {1987})}\BibitemShut {NoStop}%
\bibitem [{\citenamefont {Hamasaki}\ \emph {et~al.}(1975)\citenamefont
  {Hamasaki}, \citenamefont {Hashimoto}, \citenamefont {Yamaguchi},\ and\
  \citenamefont {Watanabe}}]{bib17}%
  \BibitemOpen
  \bibfield  {author} {\bibinfo {author} {\bibfnamefont {T.}~\bibnamefont
  {Hamasaki}}, \bibinfo {author} {\bibfnamefont {T.}~\bibnamefont {Hashimoto}},
  \bibinfo {author} {\bibfnamefont {Y.}~\bibnamefont {Yamaguchi}},\ and\
  \bibinfo {author} {\bibfnamefont {H.}~\bibnamefont {Watanabe}},\ }\bibfield
  {title} {\bibinfo {title} {{Neutron diffraction study of Cr$_2$Te$_3$ single
  crystal}},\ }\href {https://doi.org/10.1016/0038-1098(75)90888-1} {\bibfield
  {journal} {\bibinfo  {journal} {Solid State Communications}\ }\textbf
  {\bibinfo {volume} {16}},\ \bibinfo {pages} {895} (\bibinfo {year}
  {1975})}\BibitemShut {NoStop}%
\bibitem [{\citenamefont {Freitas}\ \emph {et~al.}(2015)\citenamefont
  {Freitas}, \citenamefont {Weht}, \citenamefont {Sulpice}, \citenamefont
  {Remenyi}, \citenamefont {Strobel}, \citenamefont {Gay}, \citenamefont
  {Marcus},\ and\ \citenamefont {N{\'{u}}{\~{n}}ez-Regueiro}}]{bib18}%
  \BibitemOpen
  \bibfield  {author} {\bibinfo {author} {\bibfnamefont {D.~C.}\ \bibnamefont
  {Freitas}}, \bibinfo {author} {\bibfnamefont {R.}~\bibnamefont {Weht}},
  \bibinfo {author} {\bibfnamefont {A.}~\bibnamefont {Sulpice}}, \bibinfo
  {author} {\bibfnamefont {G.}~\bibnamefont {Remenyi}}, \bibinfo {author}
  {\bibfnamefont {P.}~\bibnamefont {Strobel}}, \bibinfo {author} {\bibfnamefont
  {F.}~\bibnamefont {Gay}}, \bibinfo {author} {\bibfnamefont {J.}~\bibnamefont
  {Marcus}},\ and\ \bibinfo {author} {\bibfnamefont {M.}~\bibnamefont
  {N{\'{u}}{\~{n}}ez-Regueiro}},\ }\bibfield  {title} {\bibinfo {title}
  {{Ferromagnetism in layered metastable 1$T$-CrTe$_2$}},\ }\href
  {https://doi.org/10.1088/0953-8984/27/17/176002} {\bibfield  {journal}
  {\bibinfo  {journal} {Journal of Physics: Condensed Matter}\ }\textbf
  {\bibinfo {volume} {27}},\ \bibinfo {pages} {176002} (\bibinfo {year}
  {2015})}\BibitemShut {NoStop}%
\bibitem [{\citenamefont {Andresen}(1970)}]{bib20}%
  \BibitemOpen
  \bibfield  {author} {\bibinfo {author} {\bibfnamefont {A.~F.}\ \bibnamefont
  {Andresen}},\ }\bibfield  {title} {\bibinfo {title} {{The Magnetic Structure
  of Cr$_2$Te$_3$, Cr$_3$Te$_4$, and Cr$_5$Te$_6$}},\ }\href
  {https://doi.org/10.3891/acta.chem.scand.24-3495} {\bibfield  {journal}
  {\bibinfo  {journal} {Acta Chemica Scandinavica}\ }\textbf {\bibinfo {volume}
  {24}},\ \bibinfo {pages} {3495} (\bibinfo {year} {1970})}\BibitemShut
  {NoStop}%
\bibitem [{\citenamefont {Bester}\ \emph {et~al.}(2015)\citenamefont {Bester},
  \citenamefont {Stefaniuk},\ and\ \citenamefont {Kuzma}}]{bib19}%
  \BibitemOpen
  \bibfield  {author} {\bibinfo {author} {\bibfnamefont {M.}~\bibnamefont
  {Bester}}, \bibinfo {author} {\bibfnamefont {I.}~\bibnamefont {Stefaniuk}},\
  and\ \bibinfo {author} {\bibfnamefont {M.}~\bibnamefont {Kuzma}},\ }\bibfield
   {title} {\bibinfo {title} {{Quasi-Two-Dimensional Ferromagnetism in
  Cr$_2$Te$_3$ and Cr$_5$Te$_8$ Crystals}},\ }\href
  {https://doi.org/10.12693/APhysPolA.127.433} {\bibfield  {journal} {\bibinfo
  {journal} {Acta Physica Polonica A}\ }\textbf {\bibinfo {volume} {127}},\
  \bibinfo {pages} {433} (\bibinfo {year} {2015})}\BibitemShut {NoStop}%
\bibitem [{\citenamefont {Roy}\ \emph {et~al.}(2015)\citenamefont {Roy},
  \citenamefont {Guchhait}, \citenamefont {Dey}, \citenamefont {Pramanik},
  \citenamefont {Hsieh}, \citenamefont {Rai},\ and\ \citenamefont
  {Banerjee}}]{bib21}%
  \BibitemOpen
  \bibfield  {author} {\bibinfo {author} {\bibfnamefont {A.}~\bibnamefont
  {Roy}}, \bibinfo {author} {\bibfnamefont {S.}~\bibnamefont {Guchhait}},
  \bibinfo {author} {\bibfnamefont {R.}~\bibnamefont {Dey}}, \bibinfo {author}
  {\bibfnamefont {T.}~\bibnamefont {Pramanik}}, \bibinfo {author}
  {\bibfnamefont {C.-C.}\ \bibnamefont {Hsieh}}, \bibinfo {author}
  {\bibfnamefont {A.}~\bibnamefont {Rai}},\ and\ \bibinfo {author}
  {\bibfnamefont {S.~K.}\ \bibnamefont {Banerjee}},\ }\bibfield  {title}
  {\bibinfo {title} {{Perpendicular Magnetic Anisotropy and Spin Glass-like
  Behavior in Molecular Beam Epitaxy Grown Chromium Telluride Thin Films}},\
  }\href {https://doi.org/10.1021/nn5065716} {\bibfield  {journal} {\bibinfo
  {journal} {ACS Nano}\ }\textbf {\bibinfo {volume} {9}},\ \bibinfo {pages}
  {3772} (\bibinfo {year} {2015})}\BibitemShut {NoStop}%
\bibitem [{\citenamefont {Li}\ \emph {et~al.}(2019)\citenamefont {Li},
  \citenamefont {Wang}, \citenamefont {Chen}, \citenamefont {Yu}, \citenamefont
  {Zhou}, \citenamefont {Qiu}, \citenamefont {He}, \citenamefont {Ye},
  \citenamefont {Sou},\ and\ \citenamefont {Wang}}]{bib22}%
  \BibitemOpen
  \bibfield  {author} {\bibinfo {author} {\bibfnamefont {H.}~\bibnamefont
  {Li}}, \bibinfo {author} {\bibfnamefont {L.}~\bibnamefont {Wang}}, \bibinfo
  {author} {\bibfnamefont {J.}~\bibnamefont {Chen}}, \bibinfo {author}
  {\bibfnamefont {T.}~\bibnamefont {Yu}}, \bibinfo {author} {\bibfnamefont
  {L.}~\bibnamefont {Zhou}}, \bibinfo {author} {\bibfnamefont {Y.}~\bibnamefont
  {Qiu}}, \bibinfo {author} {\bibfnamefont {H.}~\bibnamefont {He}}, \bibinfo
  {author} {\bibfnamefont {F.}~\bibnamefont {Ye}}, \bibinfo {author}
  {\bibfnamefont {I.~K.}\ \bibnamefont {Sou}},\ and\ \bibinfo {author}
  {\bibfnamefont {G.}~\bibnamefont {Wang}},\ }\bibfield  {title} {\bibinfo
  {title} {{Molecular Beam Epitaxy Grown Cr$_2$Te$_3$ Thin Films with Tunable
  Curie Temperatures for Spintronic Devices}},\ }\href
  {https://doi.org/10.1021/acsanm.9b01179} {\bibfield  {journal} {\bibinfo
  {journal} {ACS Applied Nano Materials}\ }\textbf {\bibinfo {volume} {2}},\
  \bibinfo {pages} {6809} (\bibinfo {year} {2019})}\BibitemShut {NoStop}%
\bibitem [{\citenamefont {Coughlin}\ \emph {et~al.}(2020)\citenamefont
  {Coughlin}, \citenamefont {Xie}, \citenamefont {Yao}, \citenamefont {Zhan},
  \citenamefont {Chen}, \citenamefont {Hewa-Walpitage}, \citenamefont {Zhang},
  \citenamefont {Guo}, \citenamefont {Zhou}, \citenamefont {Lou}, \citenamefont
  {Wang}, \citenamefont {Li}, \citenamefont {Fertig},\ and\ \citenamefont
  {Zhang}}]{bib23}%
  \BibitemOpen
  \bibfield  {author} {\bibinfo {author} {\bibfnamefont {A.~L.}\ \bibnamefont
  {Coughlin}}, \bibinfo {author} {\bibfnamefont {D.}~\bibnamefont {Xie}},
  \bibinfo {author} {\bibfnamefont {Y.}~\bibnamefont {Yao}}, \bibinfo {author}
  {\bibfnamefont {X.}~\bibnamefont {Zhan}}, \bibinfo {author} {\bibfnamefont
  {Q.}~\bibnamefont {Chen}}, \bibinfo {author} {\bibfnamefont {H.}~\bibnamefont
  {Hewa-Walpitage}}, \bibinfo {author} {\bibfnamefont {X.}~\bibnamefont
  {Zhang}}, \bibinfo {author} {\bibfnamefont {H.}~\bibnamefont {Guo}}, \bibinfo
  {author} {\bibfnamefont {H.}~\bibnamefont {Zhou}}, \bibinfo {author}
  {\bibfnamefont {J.}~\bibnamefont {Lou}}, \bibinfo {author} {\bibfnamefont
  {J.}~\bibnamefont {Wang}}, \bibinfo {author} {\bibfnamefont {Y.~S.}\
  \bibnamefont {Li}}, \bibinfo {author} {\bibfnamefont {H.~A.}\ \bibnamefont
  {Fertig}},\ and\ \bibinfo {author} {\bibfnamefont {S.}~\bibnamefont
  {Zhang}},\ }\bibfield  {title} {\bibinfo {title} {{Near Degeneracy of
  Magnetic Phases in Two-Dimensional Chromium Telluride with Enhanced
  Perpendicular Magnetic Anisotropy}},\ }\href
  {https://doi.org/10.1021/acsnano.0c05534} {\bibfield  {journal} {\bibinfo
  {journal} {ACS Nano}\ }\textbf {\bibinfo {volume} {14}},\ \bibinfo {pages}
  {15256} (\bibinfo {year} {2020})}\BibitemShut {NoStop}%
\bibitem [{\citenamefont {Lee}\ \emph {et~al.}(2021)\citenamefont {Lee},
  \citenamefont {Choi}, \citenamefont {Kim}, \citenamefont {Kim}, \citenamefont
  {Jeong}, \citenamefont {Lee}, \citenamefont {Park}, \citenamefont {Jo},
  \citenamefont {Lee}, \citenamefont {Choi}, \citenamefont {Cho}, \citenamefont
  {Lee}, \citenamefont {Kim}, \citenamefont {Kim}, \citenamefont {Lee},
  \citenamefont {Heo}, \citenamefont {Chang}, \citenamefont {Li}, \citenamefont
  {Chittari}, \citenamefont {Jung},\ and\ \citenamefont {Chang}}]{bib24}%
  \BibitemOpen
  \bibfield  {author} {\bibinfo {author} {\bibfnamefont {I.~H.}\ \bibnamefont
  {Lee}}, \bibinfo {author} {\bibfnamefont {B.~K.}\ \bibnamefont {Choi}},
  \bibinfo {author} {\bibfnamefont {H.~J.}\ \bibnamefont {Kim}}, \bibinfo
  {author} {\bibfnamefont {M.~J.}\ \bibnamefont {Kim}}, \bibinfo {author}
  {\bibfnamefont {H.~Y.}\ \bibnamefont {Jeong}}, \bibinfo {author}
  {\bibfnamefont {J.~H.}\ \bibnamefont {Lee}}, \bibinfo {author} {\bibfnamefont
  {S.-Y.}\ \bibnamefont {Park}}, \bibinfo {author} {\bibfnamefont
  {Y.}~\bibnamefont {Jo}}, \bibinfo {author} {\bibfnamefont {C.}~\bibnamefont
  {Lee}}, \bibinfo {author} {\bibfnamefont {J.~W.}\ \bibnamefont {Choi}},
  \bibinfo {author} {\bibfnamefont {S.~W.}\ \bibnamefont {Cho}}, \bibinfo
  {author} {\bibfnamefont {S.}~\bibnamefont {Lee}}, \bibinfo {author}
  {\bibfnamefont {Y.}~\bibnamefont {Kim}}, \bibinfo {author} {\bibfnamefont
  {B.~H.}\ \bibnamefont {Kim}}, \bibinfo {author} {\bibfnamefont {K.~J.}\
  \bibnamefont {Lee}}, \bibinfo {author} {\bibfnamefont {J.~E.}\ \bibnamefont
  {Heo}}, \bibinfo {author} {\bibfnamefont {S.~H.}\ \bibnamefont {Chang}},
  \bibinfo {author} {\bibfnamefont {F.}~\bibnamefont {Li}}, \bibinfo {author}
  {\bibfnamefont {B.~L.}\ \bibnamefont {Chittari}}, \bibinfo {author}
  {\bibfnamefont {J.}~\bibnamefont {Jung}},\ and\ \bibinfo {author}
  {\bibfnamefont {Y.~J.}\ \bibnamefont {Chang}},\ }\bibfield  {title} {\bibinfo
  {title} {{Modulating Curie Temperature and Magnetic Anisotropy in
  Nanoscale-Layered Cr$_2$Te$_3$ Films: Implications for Room-Temperature
  Spintronics}},\ }\href {https://doi.org/10.1021/acsanm.1c00391} {\bibfield
  {journal} {\bibinfo  {journal} {ACS Applied Nano Materials}\ }\textbf
  {\bibinfo {volume} {4}},\ \bibinfo {pages} {4810} (\bibinfo {year}
  {2021})}\BibitemShut {NoStop}%
\bibitem [{\citenamefont {Fujisawa}\ \emph {et~al.}(2022)\citenamefont
  {Fujisawa}, \citenamefont {Pardo-Almanza}, \citenamefont {Hsu}, \citenamefont
  {Mohamed}, \citenamefont {Yamagami}, \citenamefont {Krishnadas},
  \citenamefont {Chuang}, \citenamefont {Khoo}, \citenamefont {Zang},
  \citenamefont {Soumyanarayanan},\ and\ \citenamefont {Okada}}]{wideBerry}%
  \BibitemOpen
  \bibfield  {author} {\bibinfo {author} {\bibfnamefont {Y.}~\bibnamefont
  {Fujisawa}}, \bibinfo {author} {\bibfnamefont {M.}~\bibnamefont
  {Pardo-Almanza}}, \bibinfo {author} {\bibfnamefont {C.~H.}\ \bibnamefont
  {Hsu}}, \bibinfo {author} {\bibfnamefont {A.}~\bibnamefont {Mohamed}},
  \bibinfo {author} {\bibfnamefont {K.}~\bibnamefont {Yamagami}}, \bibinfo
  {author} {\bibfnamefont {A.}~\bibnamefont {Krishnadas}}, \bibinfo {author}
  {\bibfnamefont {F.~C.}\ \bibnamefont {Chuang}}, \bibinfo {author}
  {\bibfnamefont {K.~H.}\ \bibnamefont {Khoo}}, \bibinfo {author}
  {\bibfnamefont {J.}~\bibnamefont {Zang}}, \bibinfo {author} {\bibfnamefont
  {A.}~\bibnamefont {Soumyanarayanan}},\ and\ \bibinfo {author} {\bibfnamefont
  {Y.}~\bibnamefont {Okada}},\ }\bibfield  {title} {\bibinfo {title} {{Widely
  Tunable Berry curvature in the Magnetic Semimetal Cr$_{1+\delta}$Te$_2$}},\
  }\href {https://doi.org/10.48550/arXiv.2204.02518} {\bibfield  {journal}
  {\bibinfo  {journal} {arXiv:2204.02518}\ } (\bibinfo {year}
  {2022})}\BibitemShut {NoStop}%
\bibitem [{\citenamefont {Zhao}\ \emph {et~al.}(2018)\citenamefont {Zhao},
  \citenamefont {Zhang}, \citenamefont {Malik}, \citenamefont {Liao},
  \citenamefont {Cui}, \citenamefont {Cai}, \citenamefont {Zheng},
  \citenamefont {Li}, \citenamefont {Hu}, \citenamefont {Zhang}, \citenamefont
  {Zhang}, \citenamefont {Chen}, \citenamefont {Jiang},\ and\ \citenamefont
  {Xue}}]{bib28}%
  \BibitemOpen
  \bibfield  {author} {\bibinfo {author} {\bibfnamefont {D.}~\bibnamefont
  {Zhao}}, \bibinfo {author} {\bibfnamefont {L.}~\bibnamefont {Zhang}},
  \bibinfo {author} {\bibfnamefont {I.~A.}\ \bibnamefont {Malik}}, \bibinfo
  {author} {\bibfnamefont {M.}~\bibnamefont {Liao}}, \bibinfo {author}
  {\bibfnamefont {W.}~\bibnamefont {Cui}}, \bibinfo {author} {\bibfnamefont
  {X.}~\bibnamefont {Cai}}, \bibinfo {author} {\bibfnamefont {C.}~\bibnamefont
  {Zheng}}, \bibinfo {author} {\bibfnamefont {L.}~\bibnamefont {Li}}, \bibinfo
  {author} {\bibfnamefont {X.}~\bibnamefont {Hu}}, \bibinfo {author}
  {\bibfnamefont {D.}~\bibnamefont {Zhang}}, \bibinfo {author} {\bibfnamefont
  {J.}~\bibnamefont {Zhang}}, \bibinfo {author} {\bibfnamefont
  {X.}~\bibnamefont {Chen}}, \bibinfo {author} {\bibfnamefont {W.}~\bibnamefont
  {Jiang}},\ and\ \bibinfo {author} {\bibfnamefont {Q.}~\bibnamefont {Xue}},\
  }\bibfield  {title} {\bibinfo {title} {{Observation of unconventional
  anomalous Hall effect in epitaxial CrTe thin films}},\ }\href
  {https://doi.org/10.1007/s12274-017-1913-8} {\bibfield  {journal} {\bibinfo
  {journal} {Nano Research}\ }\textbf {\bibinfo {volume} {11}},\ \bibinfo
  {pages} {3116} (\bibinfo {year} {2018})}\BibitemShut {NoStop}%
\bibitem [{\citenamefont {Chen}\ \emph {et~al.}(2019)\citenamefont {Chen},
  \citenamefont {Wang}, \citenamefont {Zhang}, \citenamefont {Zhou},
  \citenamefont {Zhang}, \citenamefont {Jin}, \citenamefont {Wang},
  \citenamefont {Qin}, \citenamefont {Qiu}, \citenamefont {Mei}, \citenamefont
  {Ye}, \citenamefont {Xi}, \citenamefont {He}, \citenamefont {Li},\ and\
  \citenamefont {Wang}}]{bib30}%
  \BibitemOpen
  \bibfield  {author} {\bibinfo {author} {\bibfnamefont {J.}~\bibnamefont
  {Chen}}, \bibinfo {author} {\bibfnamefont {L.}~\bibnamefont {Wang}}, \bibinfo
  {author} {\bibfnamefont {M.}~\bibnamefont {Zhang}}, \bibinfo {author}
  {\bibfnamefont {L.}~\bibnamefont {Zhou}}, \bibinfo {author} {\bibfnamefont
  {R.}~\bibnamefont {Zhang}}, \bibinfo {author} {\bibfnamefont
  {L.}~\bibnamefont {Jin}}, \bibinfo {author} {\bibfnamefont {X.}~\bibnamefont
  {Wang}}, \bibinfo {author} {\bibfnamefont {H.}~\bibnamefont {Qin}}, \bibinfo
  {author} {\bibfnamefont {Y.}~\bibnamefont {Qiu}}, \bibinfo {author}
  {\bibfnamefont {J.}~\bibnamefont {Mei}}, \bibinfo {author} {\bibfnamefont
  {F.}~\bibnamefont {Ye}}, \bibinfo {author} {\bibfnamefont {B.}~\bibnamefont
  {Xi}}, \bibinfo {author} {\bibfnamefont {H.}~\bibnamefont {He}}, \bibinfo
  {author} {\bibfnamefont {B.}~\bibnamefont {Li}},\ and\ \bibinfo {author}
  {\bibfnamefont {G.}~\bibnamefont {Wang}},\ }\bibfield  {title} {\bibinfo
  {title} {{Evidence for Magnetic Skyrmions at the Interface of
  Ferromagnet/Topological-Insulator Heterostructures}},\ }\href
  {https://doi.org/10.1021/acs.nanolett.9b02191} {\bibfield  {journal}
  {\bibinfo  {journal} {Nano Letters}\ }\textbf {\bibinfo {volume} {19}},\
  \bibinfo {pages} {6144} (\bibinfo {year} {2019})}\BibitemShut {NoStop}%
\bibitem [{\citenamefont {Zhou}\ \emph {et~al.}(2020)\citenamefont {Zhou},
  \citenamefont {Chen}, \citenamefont {Chen}, \citenamefont {Xi}, \citenamefont
  {Qiu}, \citenamefont {Zhang}, \citenamefont {Wang}, \citenamefont {Zhang},
  \citenamefont {Ye}, \citenamefont {Chen}, \citenamefont {Zhang},
  \citenamefont {Guo}, \citenamefont {Yu}, \citenamefont {Mei}, \citenamefont
  {Ye}, \citenamefont {Wang},\ and\ \citenamefont {He}}]{bib29}%
  \BibitemOpen
  \bibfield  {author} {\bibinfo {author} {\bibfnamefont {L.}~\bibnamefont
  {Zhou}}, \bibinfo {author} {\bibfnamefont {J.}~\bibnamefont {Chen}}, \bibinfo
  {author} {\bibfnamefont {X.}~\bibnamefont {Chen}}, \bibinfo {author}
  {\bibfnamefont {B.}~\bibnamefont {Xi}}, \bibinfo {author} {\bibfnamefont
  {Y.}~\bibnamefont {Qiu}}, \bibinfo {author} {\bibfnamefont {J.}~\bibnamefont
  {Zhang}}, \bibinfo {author} {\bibfnamefont {L.}~\bibnamefont {Wang}},
  \bibinfo {author} {\bibfnamefont {R.}~\bibnamefont {Zhang}}, \bibinfo
  {author} {\bibfnamefont {B.}~\bibnamefont {Ye}}, \bibinfo {author}
  {\bibfnamefont {P.}~\bibnamefont {Chen}}, \bibinfo {author} {\bibfnamefont
  {X.}~\bibnamefont {Zhang}}, \bibinfo {author} {\bibfnamefont
  {G.}~\bibnamefont {Guo}}, \bibinfo {author} {\bibfnamefont {D.}~\bibnamefont
  {Yu}}, \bibinfo {author} {\bibfnamefont {J.-W.}\ \bibnamefont {Mei}},
  \bibinfo {author} {\bibfnamefont {F.}~\bibnamefont {Ye}}, \bibinfo {author}
  {\bibfnamefont {G.}~\bibnamefont {Wang}},\ and\ \bibinfo {author}
  {\bibfnamefont {H.}~\bibnamefont {He}},\ }\bibfield  {title} {\bibinfo
  {title} {{Topological Hall Effect in Traditional Ferromagnet Embedded with
  Black-Phosphorus-Like Bismuth Nanosheets}},\ }\href
  {https://doi.org/10.1021/acsami.0c04447} {\bibfield  {journal} {\bibinfo
  {journal} {ACS Applied Materials {\&} Interfaces}\ }\textbf {\bibinfo
  {volume} {12}},\ \bibinfo {pages} {25135} (\bibinfo {year}
  {2020})}\BibitemShut {NoStop}%
\bibitem [{\citenamefont {Zhang}\ \emph
  {et~al.}(2021{\natexlab{b}})\citenamefont {Zhang}, \citenamefont {Ambhire},
  \citenamefont {Lu}, \citenamefont {Niu}, \citenamefont {Cook}, \citenamefont
  {Jiang}, \citenamefont {Hong}, \citenamefont {Alahmed}, \citenamefont {He},
  \citenamefont {Zhang}, \citenamefont {Xu}, \citenamefont {Zhang},
  \citenamefont {Li},\ and\ \citenamefont {Bian}}]{Zhang-crte2-bi2te3}%
  \BibitemOpen
  \bibfield  {author} {\bibinfo {author} {\bibfnamefont {X.}~\bibnamefont
  {Zhang}}, \bibinfo {author} {\bibfnamefont {S.~C.}\ \bibnamefont {Ambhire}},
  \bibinfo {author} {\bibfnamefont {Q.}~\bibnamefont {Lu}}, \bibinfo {author}
  {\bibfnamefont {W.}~\bibnamefont {Niu}}, \bibinfo {author} {\bibfnamefont
  {J.}~\bibnamefont {Cook}}, \bibinfo {author} {\bibfnamefont {J.~S.}\
  \bibnamefont {Jiang}}, \bibinfo {author} {\bibfnamefont {D.}~\bibnamefont
  {Hong}}, \bibinfo {author} {\bibfnamefont {L.}~\bibnamefont {Alahmed}},
  \bibinfo {author} {\bibfnamefont {L.}~\bibnamefont {He}}, \bibinfo {author}
  {\bibfnamefont {R.}~\bibnamefont {Zhang}}, \bibinfo {author} {\bibfnamefont
  {Y.}~\bibnamefont {Xu}}, \bibinfo {author} {\bibfnamefont {S.~S.~L.}\
  \bibnamefont {Zhang}}, \bibinfo {author} {\bibfnamefont {P.}~\bibnamefont
  {Li}},\ and\ \bibinfo {author} {\bibfnamefont {G.}~\bibnamefont {Bian}},\
  }\bibfield  {title} {\bibinfo {title} {{Giant Topological Hall Effect in van
  der Waals Heterostructures of CrTe$_2$/Bi$_2$Te$_3$}},\ }\href
  {https://doi.org/10.1021/acsnano.1c05519} {\bibfield  {journal} {\bibinfo
  {journal} {ACS Nano}\ }\textbf {\bibinfo {volume} {15}},\ \bibinfo {pages}
  {15710} (\bibinfo {year} {2021}{\natexlab{b}})}\BibitemShut {NoStop}%
\bibitem [{\citenamefont {Jeon}\ \emph {et~al.}(2022)\citenamefont {Jeon},
  \citenamefont {Na}, \citenamefont {Kim}, \citenamefont {Lee}, \citenamefont
  {Song}, \citenamefont {Kim}, \citenamefont {Park}, \citenamefont {Kim},
  \citenamefont {Noh}, \citenamefont {Kim}, \citenamefont {Jerng},\ and\
  \citenamefont {Chun}}]{Jeon-cr2te3-cr2se3}%
  \BibitemOpen
  \bibfield  {author} {\bibinfo {author} {\bibfnamefont {J.~H.}\ \bibnamefont
  {Jeon}}, \bibinfo {author} {\bibfnamefont {H.~R.}\ \bibnamefont {Na}},
  \bibinfo {author} {\bibfnamefont {H.}~\bibnamefont {Kim}}, \bibinfo {author}
  {\bibfnamefont {S.}~\bibnamefont {Lee}}, \bibinfo {author} {\bibfnamefont
  {S.}~\bibnamefont {Song}}, \bibinfo {author} {\bibfnamefont {J.}~\bibnamefont
  {Kim}}, \bibinfo {author} {\bibfnamefont {S.}~\bibnamefont {Park}}, \bibinfo
  {author} {\bibfnamefont {J.}~\bibnamefont {Kim}}, \bibinfo {author}
  {\bibfnamefont {H.}~\bibnamefont {Noh}}, \bibinfo {author} {\bibfnamefont
  {G.}~\bibnamefont {Kim}}, \bibinfo {author} {\bibfnamefont {S.-K.}\
  \bibnamefont {Jerng}},\ and\ \bibinfo {author} {\bibfnamefont {S.-H.}\
  \bibnamefont {Chun}},\ }\bibfield  {title} {\bibinfo {title} {{Emergent
  Topological Hall Effect from Exchange Coupling in Ferromagnetic
  Cr$_2$Te$_3$/Noncoplanar Antiferromagnetic Cr$_2$Se$_3$ Bilayers}},\ }\href
  {https://doi.org/10.1021/acsnano.2c00025} {\bibfield  {journal} {\bibinfo
  {journal} {ACS Nano}\ }\textbf {\bibinfo {volume} {16}},\ \bibinfo {pages}
  {8974} (\bibinfo {year} {2022})}\BibitemShut {NoStop}%
\bibitem [{\citenamefont {Ou}\ \emph {et~al.}(2022)\citenamefont {Ou},
  \citenamefont {Yanez}, \citenamefont {Xiao}, \citenamefont {Stanley},
  \citenamefont {Ghosh}, \citenamefont {Zheng}, \citenamefont {Jiang},
  \citenamefont {Huang}, \citenamefont {Pillsbury}, \citenamefont
  {Richardella}, \citenamefont {Liu}, \citenamefont {Low}, \citenamefont
  {Crespi}, \citenamefont {Mkhoyan},\ and\ \citenamefont
  {Samarth}}]{Ou-crte2-zrte2}%
  \BibitemOpen
  \bibfield  {author} {\bibinfo {author} {\bibfnamefont {Y.}~\bibnamefont
  {Ou}}, \bibinfo {author} {\bibfnamefont {W.}~\bibnamefont {Yanez}}, \bibinfo
  {author} {\bibfnamefont {R.}~\bibnamefont {Xiao}}, \bibinfo {author}
  {\bibfnamefont {M.}~\bibnamefont {Stanley}}, \bibinfo {author} {\bibfnamefont
  {S.}~\bibnamefont {Ghosh}}, \bibinfo {author} {\bibfnamefont
  {B.}~\bibnamefont {Zheng}}, \bibinfo {author} {\bibfnamefont
  {W.}~\bibnamefont {Jiang}}, \bibinfo {author} {\bibfnamefont {Y.-S.}\
  \bibnamefont {Huang}}, \bibinfo {author} {\bibfnamefont {T.}~\bibnamefont
  {Pillsbury}}, \bibinfo {author} {\bibfnamefont {A.}~\bibnamefont
  {Richardella}}, \bibinfo {author} {\bibfnamefont {C.}~\bibnamefont {Liu}},
  \bibinfo {author} {\bibfnamefont {T.}~\bibnamefont {Low}}, \bibinfo {author}
  {\bibfnamefont {V.~H.}\ \bibnamefont {Crespi}}, \bibinfo {author}
  {\bibfnamefont {K.~A.}\ \bibnamefont {Mkhoyan}},\ and\ \bibinfo {author}
  {\bibfnamefont {N.}~\bibnamefont {Samarth}},\ }\bibfield  {title} {\bibinfo
  {title} {{ZrTe$_2$/CrTe$_2$: an epitaxial van der Waals platform for
  spintronics}},\ }\href {https://doi.org/10.1038/s41467-022-30738-1}
  {\bibfield  {journal} {\bibinfo  {journal} {Nature Communications}\ }\textbf
  {\bibinfo {volume} {13}},\ \bibinfo {pages} {2972} (\bibinfo {year}
  {2022})}\BibitemShut {NoStop}%
\bibitem [{\citenamefont {Kimbell}\ \emph {et~al.}(2022)\citenamefont
  {Kimbell}, \citenamefont {Kim}, \citenamefont {Wu}, \citenamefont {Cuoco},\
  and\ \citenamefont {Robinson}}]{Kimbell-2022-CM}%
  \BibitemOpen
  \bibfield  {author} {\bibinfo {author} {\bibfnamefont {G.}~\bibnamefont
  {Kimbell}}, \bibinfo {author} {\bibfnamefont {C.}~\bibnamefont {Kim}},
  \bibinfo {author} {\bibfnamefont {W.}~\bibnamefont {Wu}}, \bibinfo {author}
  {\bibfnamefont {M.}~\bibnamefont {Cuoco}},\ and\ \bibinfo {author}
  {\bibfnamefont {J.~W.~A.}\ \bibnamefont {Robinson}},\ }\bibfield  {title}
  {\bibinfo {title} {{Challenges in identifying chiral spin textures via the
  topological Hall effect}},\ }\href
  {https://doi.org/10.1038/s43246-022-00238-2} {\bibfield  {journal} {\bibinfo
  {journal} {Communications Materials}\ }\textbf {\bibinfo {volume} {3}},\
  \bibinfo {pages} {19} (\bibinfo {year} {2022})}\BibitemShut {NoStop}%
\bibitem [{\citenamefont {Kan}\ \emph {et~al.}(2018)\citenamefont {Kan},
  \citenamefont {Moriyama}, \citenamefont {Kobayashi},\ and\ \citenamefont
  {Shimakawa}}]{bib27}%
  \BibitemOpen
  \bibfield  {author} {\bibinfo {author} {\bibfnamefont {D.}~\bibnamefont
  {Kan}}, \bibinfo {author} {\bibfnamefont {T.}~\bibnamefont {Moriyama}},
  \bibinfo {author} {\bibfnamefont {K.}~\bibnamefont {Kobayashi}},\ and\
  \bibinfo {author} {\bibfnamefont {Y.}~\bibnamefont {Shimakawa}},\ }\bibfield
  {title} {\bibinfo {title} {{Alternative to the topological interpretation of
  the transverse resistivity anomalies in SrRuO$_3$}},\ }\href
  {https://doi.org/10.1103/PhysRevB.98.180408} {\bibfield  {journal} {\bibinfo
  {journal} {Physical Review B}\ }\textbf {\bibinfo {volume} {98}},\ \bibinfo
  {pages} {180408} (\bibinfo {year} {2018})}\BibitemShut {NoStop}%
\bibitem [{\citenamefont {Dreyer}\ \emph {et~al.}(2010)\citenamefont {Dreyer},
  \citenamefont {Lee}, \citenamefont {Wang},\ and\ \citenamefont
  {Barker}}]{dreyer-2010-rsi}%
  \BibitemOpen
  \bibfield  {author} {\bibinfo {author} {\bibfnamefont {M.}~\bibnamefont
  {Dreyer}}, \bibinfo {author} {\bibfnamefont {J.}~\bibnamefont {Lee}},
  \bibinfo {author} {\bibfnamefont {H.}~\bibnamefont {Wang}},\ and\ \bibinfo
  {author} {\bibfnamefont {B.}~\bibnamefont {Barker}},\ }\bibfield  {title}
  {\bibinfo {title} {{A low temperature scanning tunneling microscopy system
  for measuring Si at 4.2 K}},\ }\href {https://doi.org/10.1063/1.3427217}
  {\bibfield  {journal} {\bibinfo  {journal} {Review of Scientific
  Instruments}\ }\textbf {\bibinfo {volume} {81}},\ \bibinfo {pages} {053703}
  (\bibinfo {year} {2010})}\BibitemShut {NoStop}%
\bibitem [{\citenamefont {Blundell}\ \emph {et~al.}(1995)\citenamefont
  {Blundell}, \citenamefont {Gester}, \citenamefont {Bland}, \citenamefont
  {Lauter}, \citenamefont {Pasyuk},\ and\ \citenamefont
  {Petrenko}}]{pnr-method-1}%
  \BibitemOpen
  \bibfield  {author} {\bibinfo {author} {\bibfnamefont {S.~J.}\ \bibnamefont
  {Blundell}}, \bibinfo {author} {\bibfnamefont {M.}~\bibnamefont {Gester}},
  \bibinfo {author} {\bibfnamefont {J.~A.~C.}\ \bibnamefont {Bland}}, \bibinfo
  {author} {\bibfnamefont {H.~J.}\ \bibnamefont {Lauter}}, \bibinfo {author}
  {\bibfnamefont {V.~V.}\ \bibnamefont {Pasyuk}},\ and\ \bibinfo {author}
  {\bibfnamefont {A.~V.}\ \bibnamefont {Petrenko}},\ }\bibfield  {title}
  {\bibinfo {title} {{Spin-orientation dependence in neutron reflection from a
  single magnetic film}},\ }\href {https://doi.org/10.1103/PhysRevB.51.9395}
  {\bibfield  {journal} {\bibinfo  {journal} {Physical Review B}\ }\textbf
  {\bibinfo {volume} {51}},\ \bibinfo {pages} {9395} (\bibinfo {year}
  {1995})}\BibitemShut {NoStop}%
\bibitem [{\citenamefont {Lauter-Pasyuk}\ \emph {et~al.}(2000)\citenamefont
  {Lauter-Pasyuk}, \citenamefont {Lauter}, \citenamefont {Toperverg},
  \citenamefont {Nikonov}, \citenamefont {Kravtsov}, \citenamefont {Milyaev},
  \citenamefont {Romashev},\ and\ \citenamefont {Ustinov}}]{pnr-method-2}%
  \BibitemOpen
  \bibfield  {author} {\bibinfo {author} {\bibfnamefont {V.}~\bibnamefont
  {Lauter-Pasyuk}}, \bibinfo {author} {\bibfnamefont {H.~J.}\ \bibnamefont
  {Lauter}}, \bibinfo {author} {\bibfnamefont {B.}~\bibnamefont {Toperverg}},
  \bibinfo {author} {\bibfnamefont {O.}~\bibnamefont {Nikonov}}, \bibinfo
  {author} {\bibfnamefont {E.}~\bibnamefont {Kravtsov}}, \bibinfo {author}
  {\bibfnamefont {M.~A.}\ \bibnamefont {Milyaev}}, \bibinfo {author}
  {\bibfnamefont {L.}~\bibnamefont {Romashev}},\ and\ \bibinfo {author}
  {\bibfnamefont {V.}~\bibnamefont {Ustinov}},\ }\bibfield  {title} {\bibinfo
  {title} {{Magnetic off-specular neutron scattering from Fe/Cr multilayers}},\
  }\href {https://doi.org/10.1016/S0921-4526(99)01938-9} {\bibfield  {journal}
  {\bibinfo  {journal} {Physica B: Condensed Matter}\ }\textbf {\bibinfo
  {volume} {283}},\ \bibinfo {pages} {194} (\bibinfo {year}
  {2000})}\BibitemShut {NoStop}%
\bibitem [{\citenamefont {Lauter-Pasyuk}(2007)}]{pnr-method-3}%
  \BibitemOpen
  \bibfield  {author} {\bibinfo {author} {\bibfnamefont {V.}~\bibnamefont
  {Lauter-Pasyuk}},\ }\bibfield  {title} {\bibinfo {title} {{Neutron grazing
  incidence techniques for nano-science}},\ }\href@noop {} {\bibfield
  {journal} {\bibinfo  {journal} {Collection SFN}\ }\textbf {\bibinfo {volume}
  {7}},\ \bibinfo {pages} {s221} (\bibinfo {year} {2007})}\BibitemShut
  {NoStop}%
\bibitem [{\citenamefont {Lauter}\ \emph {et~al.}(2016)\citenamefont {Lauter},
  \citenamefont {Lauter}, \citenamefont {Glavic},\ and\ \citenamefont
  {Toperverg}}]{pnr-method-4}%
  \BibitemOpen
  \bibfield  {author} {\bibinfo {author} {\bibfnamefont {V.}~\bibnamefont
  {Lauter}}, \bibinfo {author} {\bibfnamefont {H.~J.~C.}\ \bibnamefont
  {Lauter}}, \bibinfo {author} {\bibfnamefont {A.}~\bibnamefont {Glavic}},\
  and\ \bibinfo {author} {\bibfnamefont {B.~P.}\ \bibnamefont {Toperverg}},\
  }\bibinfo {title} {{Reflectivity, Off-Specular Scattering, and GISANS
  Neutrons}},\ in\ \href
  {https://doi.org/https://doi.org/10.1016/B978-0-12-803581-8.01324-2} {\emph
  {\bibinfo {booktitle} {{Reference Module in Materials Science and Materials
  Engineering}}}},\ \bibinfo {editor} {edited by\ \bibinfo {editor}
  {\bibfnamefont {S.}~\bibnamefont {Hashmi}}}\ (\bibinfo  {publisher} {Oxford:
  Elsevier},\ \bibinfo {year} {2016})\ pp.\ \bibinfo {pages}
  {1--27}\BibitemShut {NoStop}%
\bibitem [{\citenamefont {Lauter}\ \emph {et~al.}(2009)\citenamefont {Lauter},
  \citenamefont {Ambaye}, \citenamefont {Goyette}, \citenamefont {Hal~Lee},\
  and\ \citenamefont {Parizzi}}]{pnrValeria}%
  \BibitemOpen
  \bibfield  {author} {\bibinfo {author} {\bibfnamefont {V.}~\bibnamefont
  {Lauter}}, \bibinfo {author} {\bibfnamefont {H.}~\bibnamefont {Ambaye}},
  \bibinfo {author} {\bibfnamefont {R.}~\bibnamefont {Goyette}}, \bibinfo
  {author} {\bibfnamefont {W.-T.}\ \bibnamefont {Hal~Lee}},\ and\ \bibinfo
  {author} {\bibfnamefont {A.}~\bibnamefont {Parizzi}},\ }\bibfield  {title}
  {\bibinfo {title} {{Highlights from the magnetism reflectometer at the
  SNS}},\ }\href {https://doi.org/10.1016/j.physb.2009.06.021} {\bibfield
  {journal} {\bibinfo  {journal} {Physica B: Condensed Matter}\ }\textbf
  {\bibinfo {volume} {404}},\ \bibinfo {pages} {2543} (\bibinfo {year}
  {2009})}\BibitemShut {NoStop}%
\bibitem [{\citenamefont {Jiang}\ \emph {et~al.}(2017)\citenamefont {Jiang},
  \citenamefont {Tong}, \citenamefont {Brown}, \citenamefont {Glavic},
  \citenamefont {Ambaye}, \citenamefont {Goyette}, \citenamefont {Hoffmann},
  \citenamefont {Parizzi}, \citenamefont {Robertson},\ and\ \citenamefont
  {Lauter}}]{pnrJiang}%
  \BibitemOpen
  \bibfield  {author} {\bibinfo {author} {\bibfnamefont {C.~Y.}\ \bibnamefont
  {Jiang}}, \bibinfo {author} {\bibfnamefont {X.}~\bibnamefont {Tong}},
  \bibinfo {author} {\bibfnamefont {D.~R.}\ \bibnamefont {Brown}}, \bibinfo
  {author} {\bibfnamefont {A.}~\bibnamefont {Glavic}}, \bibinfo {author}
  {\bibfnamefont {H.}~\bibnamefont {Ambaye}}, \bibinfo {author} {\bibfnamefont
  {R.}~\bibnamefont {Goyette}}, \bibinfo {author} {\bibfnamefont
  {M.}~\bibnamefont {Hoffmann}}, \bibinfo {author} {\bibfnamefont {A.~A.}\
  \bibnamefont {Parizzi}}, \bibinfo {author} {\bibfnamefont {L.}~\bibnamefont
  {Robertson}},\ and\ \bibinfo {author} {\bibfnamefont {V.}~\bibnamefont
  {Lauter}},\ }\bibfield  {title} {\bibinfo {title} {{New generation high
  performance in situ polarized $^3$He system for time-of-flight beam at
  spallation sources}},\ }\href {https://doi.org/10.1063/1.4975991} {\bibfield
  {journal} {\bibinfo  {journal} {Review of Scientific Instruments}\ }\textbf
  {\bibinfo {volume} {88}},\ \bibinfo {pages} {025111} (\bibinfo {year}
  {2017})}\BibitemShut {NoStop}%
\bibitem [{\citenamefont {Syromyatnikov}\ \emph {et~al.}(2014)\citenamefont
  {Syromyatnikov}, \citenamefont {Ulyanov}, \citenamefont {Lauter},
  \citenamefont {Pusenkov}, \citenamefont {Ambaye}, \citenamefont {Goyette},
  \citenamefont {Hoffmann}, \citenamefont {Bulkin}, \citenamefont {Kuznetsov},\
  and\ \citenamefont {Medvedev}}]{pnrSyro}%
  \BibitemOpen
  \bibfield  {author} {\bibinfo {author} {\bibfnamefont {V.~G.}\ \bibnamefont
  {Syromyatnikov}}, \bibinfo {author} {\bibfnamefont {V.~A.}\ \bibnamefont
  {Ulyanov}}, \bibinfo {author} {\bibfnamefont {V.}~\bibnamefont {Lauter}},
  \bibinfo {author} {\bibfnamefont {V.~M.}\ \bibnamefont {Pusenkov}}, \bibinfo
  {author} {\bibfnamefont {H.}~\bibnamefont {Ambaye}}, \bibinfo {author}
  {\bibfnamefont {R.}~\bibnamefont {Goyette}}, \bibinfo {author} {\bibfnamefont
  {M.}~\bibnamefont {Hoffmann}}, \bibinfo {author} {\bibfnamefont {A.~P.}\
  \bibnamefont {Bulkin}}, \bibinfo {author} {\bibfnamefont {I.~N.}\
  \bibnamefont {Kuznetsov}},\ and\ \bibinfo {author} {\bibfnamefont {E.~N.}\
  \bibnamefont {Medvedev}},\ }\bibfield  {title} {\bibinfo {title} {A new type
  of wide-angle supermirror analyzer of neutron polarization},\ }\href
  {https://doi.org/10.1088/1742-6596/528/1/012021} {\bibfield  {journal}
  {\bibinfo  {journal} {Journal of Physics: Conference Series}\ }\textbf
  {\bibinfo {volume} {528}},\ \bibinfo {pages} {012021} (\bibinfo {year}
  {2014})}\BibitemShut {NoStop}%
\bibitem [{\citenamefont {Giannozzi}\ \emph {et~al.}(2009)\citenamefont
  {Giannozzi}, \citenamefont {Baroni}, \citenamefont {Bonini}, \citenamefont
  {Calandra}, \citenamefont {Car}, \citenamefont {Cavazzoni}, \citenamefont
  {Ceresoli}, \citenamefont {Chiarotti}, \citenamefont {Cococcioni},
  \citenamefont {Dabo}, \citenamefont {Dal~Corso}, \citenamefont
  {de~Gironcoli}, \citenamefont {Fabris}, \citenamefont {Fratesi},
  \citenamefont {Gebauer} \emph {et~al.}}]{dft-method-1}%
  \BibitemOpen
  \bibfield  {author} {\bibinfo {author} {\bibfnamefont {P.}~\bibnamefont
  {Giannozzi}}, \bibinfo {author} {\bibfnamefont {S.}~\bibnamefont {Baroni}},
  \bibinfo {author} {\bibfnamefont {N.}~\bibnamefont {Bonini}}, \bibinfo
  {author} {\bibfnamefont {M.}~\bibnamefont {Calandra}}, \bibinfo {author}
  {\bibfnamefont {R.}~\bibnamefont {Car}}, \bibinfo {author} {\bibfnamefont
  {C.}~\bibnamefont {Cavazzoni}}, \bibinfo {author} {\bibfnamefont
  {D.}~\bibnamefont {Ceresoli}}, \bibinfo {author} {\bibfnamefont {G.~L.}\
  \bibnamefont {Chiarotti}}, \bibinfo {author} {\bibfnamefont {M.}~\bibnamefont
  {Cococcioni}}, \bibinfo {author} {\bibfnamefont {I.}~\bibnamefont {Dabo}},
  \bibinfo {author} {\bibfnamefont {A.}~\bibnamefont {Dal~Corso}}, \bibinfo
  {author} {\bibfnamefont {S.}~\bibnamefont {de~Gironcoli}}, \bibinfo {author}
  {\bibfnamefont {S.}~\bibnamefont {Fabris}}, \bibinfo {author} {\bibfnamefont
  {G.}~\bibnamefont {Fratesi}}, \bibinfo {author} {\bibfnamefont
  {R.}~\bibnamefont {Gebauer}}, \emph {et~al.},\ }\bibfield  {title} {\bibinfo
  {title} {{QUANTUM ESPRESSO: a modular and open-source software project for
  quantum simulations of materials}},\ }\href
  {https://doi.org/10.1088/0953-8984/21/39/395502} {\bibfield  {journal}
  {\bibinfo  {journal} {Journal of Physics: Condensed Matter}\ }\textbf
  {\bibinfo {volume} {21}},\ \bibinfo {pages} {395502} (\bibinfo {year}
  {2009})}\BibitemShut {NoStop}%
\bibitem [{\citenamefont {Perdew}\ \emph {et~al.}(1996)\citenamefont {Perdew},
  \citenamefont {Burke},\ and\ \citenamefont {Ernzerhof}}]{dft-method-2}%
  \BibitemOpen
  \bibfield  {author} {\bibinfo {author} {\bibfnamefont {J.~P.}\ \bibnamefont
  {Perdew}}, \bibinfo {author} {\bibfnamefont {K.}~\bibnamefont {Burke}},\ and\
  \bibinfo {author} {\bibfnamefont {M.}~\bibnamefont {Ernzerhof}},\ }\bibfield
  {title} {\bibinfo {title} {{Generalized Gradient Approximation Made
  Simple}},\ }\href {https://doi.org/10.1103/PhysRevLett.77.3865} {\bibfield
  {journal} {\bibinfo  {journal} {Physical Review Letters}\ }\textbf {\bibinfo
  {volume} {77}},\ \bibinfo {pages} {3865} (\bibinfo {year}
  {1996})}\BibitemShut {NoStop}%
\bibitem [{\citenamefont {Pizzi}\ \emph {et~al.}(2020)\citenamefont {Pizzi},
  \citenamefont {Vitale}, \citenamefont {Arita}, \citenamefont {Blügel},
  \citenamefont {Freimuth}, \citenamefont {Géranton}, \citenamefont
  {Gibertini}, \citenamefont {Gresch}, \citenamefont {Johnson}, \citenamefont
  {Koretsune}, \citenamefont {Ibañez-Azpiroz}, \citenamefont {Lee},
  \citenamefont {Lihm}, \citenamefont {Marchand}, \citenamefont {Marrazzo}
  \emph {et~al.}}]{dft-method-3}%
  \BibitemOpen
  \bibfield  {author} {\bibinfo {author} {\bibfnamefont {G.}~\bibnamefont
  {Pizzi}}, \bibinfo {author} {\bibfnamefont {V.}~\bibnamefont {Vitale}},
  \bibinfo {author} {\bibfnamefont {R.}~\bibnamefont {Arita}}, \bibinfo
  {author} {\bibfnamefont {S.}~\bibnamefont {Blügel}}, \bibinfo {author}
  {\bibfnamefont {F.}~\bibnamefont {Freimuth}}, \bibinfo {author}
  {\bibfnamefont {G.}~\bibnamefont {Géranton}}, \bibinfo {author}
  {\bibfnamefont {M.}~\bibnamefont {Gibertini}}, \bibinfo {author}
  {\bibfnamefont {D.}~\bibnamefont {Gresch}}, \bibinfo {author} {\bibfnamefont
  {C.}~\bibnamefont {Johnson}}, \bibinfo {author} {\bibfnamefont
  {T.}~\bibnamefont {Koretsune}}, \bibinfo {author} {\bibfnamefont
  {J.}~\bibnamefont {Ibañez-Azpiroz}}, \bibinfo {author} {\bibfnamefont
  {H.}~\bibnamefont {Lee}}, \bibinfo {author} {\bibfnamefont {J.-M.}\
  \bibnamefont {Lihm}}, \bibinfo {author} {\bibfnamefont {D.}~\bibnamefont
  {Marchand}}, \bibinfo {author} {\bibfnamefont {A.}~\bibnamefont {Marrazzo}},
  \emph {et~al.},\ }\bibfield  {title} {\bibinfo {title} {Wannier90 as a
  community code: new features and applications},\ }\href
  {https://doi.org/10.1088/1361-648x/ab51ff} {\bibfield  {journal} {\bibinfo
  {journal} {Journal of Physics: Condensed Matter}\ }\textbf {\bibinfo {volume}
  {32}},\ \bibinfo {pages} {165902} (\bibinfo {year} {2020})}\BibitemShut
  {NoStop}%
\end{thebibliography}%

\end{document}

%\preprint{APS/123-QED}

%\title{Supplementary Information \linebreak 
%Strain tunable Berry curvature in quasi-two-dimensional chromium telluride}
%\author{H. Chi \textit{et al.}}    
%\date{\today}

\title{Supplementary Information for \\ Strain-tunable Berry curvature in quasi-two-dimensional chromium telluride}
\author{Hang Chi}    
%    \email{chihang@mit.edu}
    \affiliation{Francis Bitter Magnet Laboratory, Plasma Science and Fusion Center, Massachusetts Institute of Technology, Cambridge, Massachusetts 02139, USA}
    \affiliation{U.S. Army CCDC Army Research Laboratory, Adelphi, Maryland 20783, USA}
\author{Yunbo Ou}    
%    \email{ybou@mit.edu}
    \affiliation{Francis Bitter Magnet Laboratory, Plasma Science and Fusion Center, Massachusetts Institute of Technology, Cambridge, Massachusetts 02139, USA}
\author{Tim B. Eldred}    
    \affiliation{Department of Materials Science and Engineering, North Carolina State University, Raleigh, North Carolina 27695, USA}
\author{Wenpei Gao}    
    \affiliation{Department of Materials Science and Engineering, North Carolina State University, Raleigh, North Carolina 27695, USA}
\author{Sohee Kwon}    
    \affiliation{Department of Electrical and Computer Engineering, 
University of California, Riverside, California 92521, USA}
\author{Joseph Murray}    
    \affiliation{Department of Physics, University of Maryland, College Park, Maryland 20742, USA}
\author{Michael Dreyer}    
    \affiliation{Department of Physics, University of Maryland, College Park, Maryland 20742, USA}
\author{Robert E. Butera}    
    \affiliation{Laboratory for Physical Sciences, College Park, Maryland 20740, USA}
\author{Alexandre C. Foucher}    
    \affiliation{Department of Materials Science and Engineering, Massachusetts Institute of Technology, Cambridge, Massachusetts 02139, USA}
\author{Haile Ambaye}    
    \affiliation{Neutron Scattering Division, Neutron Sciences Directorate, Oak Ridge National Laboratory, Oak Ridge, Tennessee 37831, USA}
\author{Jong Keum}    
    \affiliation{Neutron Scattering Division, Neutron Sciences Directorate, Oak Ridge National Laboratory, Oak Ridge, Tennessee 37831, USA}
    \affiliation{Center for Nanophase Materials Sciences, Physical Science Directorate, Oak Ridge National Laboratory, Oak Ridge, Tennessee 37831, USA}
\author{Alice T. Greenberg}    
    \affiliation{U.S. Army CCDC Army Research Laboratory, Adelphi, Maryland 20783, USA}
\author{Yuhang Liu}    
    \affiliation{Department of Electrical and Computer Engineering, 
University of California, Riverside, California 92521, USA}
\author{Mahesh R. Neupane}    
    \affiliation{U.S. Army CCDC Army Research Laboratory, Adelphi, Maryland 20783, USA}
    \affiliation{Department of Electrical and Computer Engineering, 
University of California, Riverside, California 92521, USA}
\author{George J. de Coster}    
    \affiliation{U.S. Army CCDC Army Research Laboratory, Adelphi, Maryland 20783, USA}
\author{Owen A. Vail}    
    \affiliation{U.S. Army CCDC Army Research Laboratory, Adelphi, Maryland 20783, USA}
\author{Patrick J. Taylor}    
    \affiliation{U.S. Army CCDC Army Research Laboratory, Adelphi, Maryland 20783, USA}
\author{Patrick A. Folkes}    
    \affiliation{U.S. Army CCDC Army Research Laboratory, Adelphi, Maryland 20783, USA}
\author{Charles Rong}    
    \affiliation{U.S. Army CCDC Army Research Laboratory, Adelphi, Maryland 20783, USA}
\author{Gen Yin}    
    \affiliation{Department of Physics, Georgetown University, Washington, District of Columbia 20057, USA}
\author{Roger K. Lake}    
    \affiliation{Department of Electrical and Computer Engineering, 
University of California, Riverside, California 92521, USA}
\author{Frances M. Ross}    
    \affiliation{Department of Materials Science and Engineering, Massachusetts Institute of Technology, Cambridge, Massachusetts 02139, USA}
\author{Valeria Lauter}    
    \affiliation{Neutron Scattering Division, Neutron Sciences Directorate, Oak Ridge National Laboratory, Oak Ridge, Tennessee 37831, USA}
\author{Don Heiman}    
    \affiliation{Francis Bitter Magnet Laboratory, Plasma Science and Fusion Center, Massachusetts Institute of Technology, Cambridge, Massachusetts 02139, USA}
    \affiliation{Department of Physics, Northeastern University, Boston, Massachusetts 02115, USA}
\author{Jagadeesh S. Moodera}    
%    \email{moodera@mit.edu}
    \affiliation{Francis Bitter Magnet Laboratory, Plasma Science and Fusion Center, Massachusetts Institute of Technology, Cambridge, Massachusetts 02139, USA}
    \affiliation{Department of Physics, Massachusetts Institute of Technology, Cambridge, Massachusetts 02139, USA}

\date{\today}
%\date{July 6, 2022}
\maketitle

\onecolumngrid

%\tableofcontents

%\clearpage

%\section{\label{sec:level1}Crystal Structure}
\begin{figure*}%[!thb]
\includegraphics{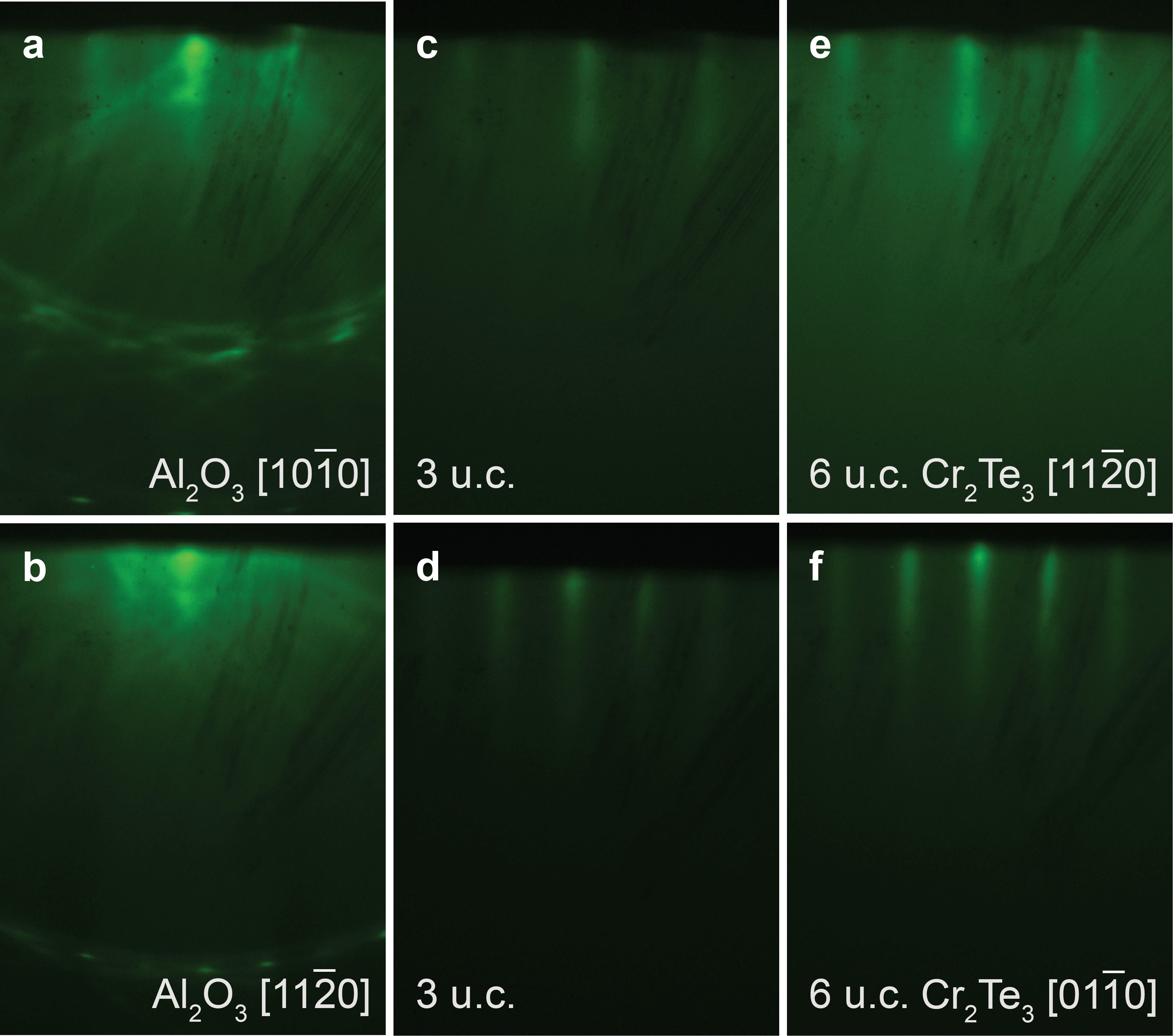}
\mycaption{\label{fig:figs1}
\textbf{Reflection high-energy electron diffraction of Cr$_2$Te$_3$.} Typical \textit{in situ} RHEED patterns from the surface of heat-treated Al$_2$O$_3$(0001) substrates (\textbf{a}, \textbf{b}) and as-grown $c$-oriented Cr$_2$Te$_3$ (\textbf{c}-\textbf{f}) with thickness of 3 (\textbf{c}, \textbf{d}) and 6 (\textbf{e}, \textbf{f}) unit cell (u.c.). The incident electron beam is along the $[10\bar{1}0]$ (\textbf{a}, \textbf{c}, \textbf{e}) and $[11\bar{2}0]$ (\textbf{b}, \textbf{d}, \textbf{f}) crystalline orientations of Al$_2$O$_3$, respectively. The clear Kikuchi lines in \textbf{a} and \textbf{b} attest to an atomically flat surface ready for the fabrication of high-quality epitaxial films. The corresponding RHEED patterns from the $c$-oriented Cr$_2$Te$_3$ (001) surface reveal sharp and streaky diffraction during the film deposition process, indicating the formation of a highly ordered and smooth surface as well as a 2D growth mode. Upon in-plane rotation, the same RHEED patterns reemerge every 60$^\circ$, suggesting a six-fold crystalline symmetry within the basal plane of the as-grown films.}
\end{figure*}

\begin{figure*}%[!hb]
\includegraphics{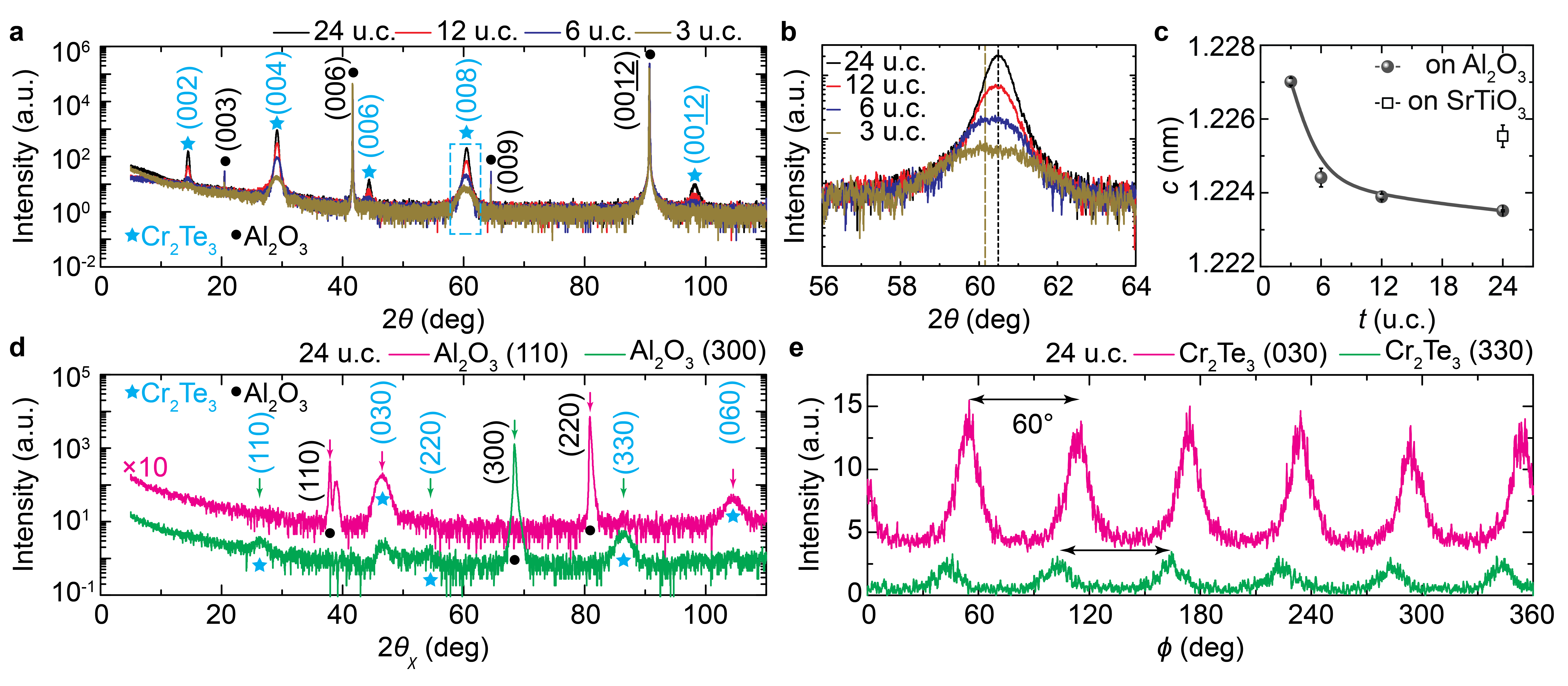}
\mycaption{\label{fig:figs2}
\textbf{X-ray diffraction of Cr$_2$Te$_3$.} \textbf{a}, Out-of-plane $2\theta/\omega$ XRD patterns of $c$-oriented Cr$_2$Te$_3$ on Al$_2$O$_3$(0001) with thickness $t$ = 3 -- 24 u.c.. \textbf{b}, Enlarged view of the (008) peaks showing gradual shift towards lower $2\theta$ upon decreasing $t$. \textbf{c}, The $t$-dependence of the $c$ lattice parameter indicating enhanced in-plane compressive strain at reduced $t$. \textbf{d}, In-plane $2\theta_{\chi}/\phi$ scans for $t$ = 24 u.c. Cr$_2$Te$_3$ thin film, aligned with the Al$_2$O$_3$ (110) and (300) orientations. \textbf{e}, X-ray $\phi$ scans with $2\theta_{\chi}$ angle fixed at Cr$_2$Te$_3$ (030) and (330), respectively, corroborating the in-plane six-fold rotational symmetry. The lattice parameters are measured to be $a$ = $b$ = 0.675 ($\pm$ 0.002) nm and $c$ = 1.223 ($\pm$ 0.004) nm, respectively, for $t$ = 24 u.c..}
\end{figure*}

\begin{figure*}%[htb]
\includegraphics{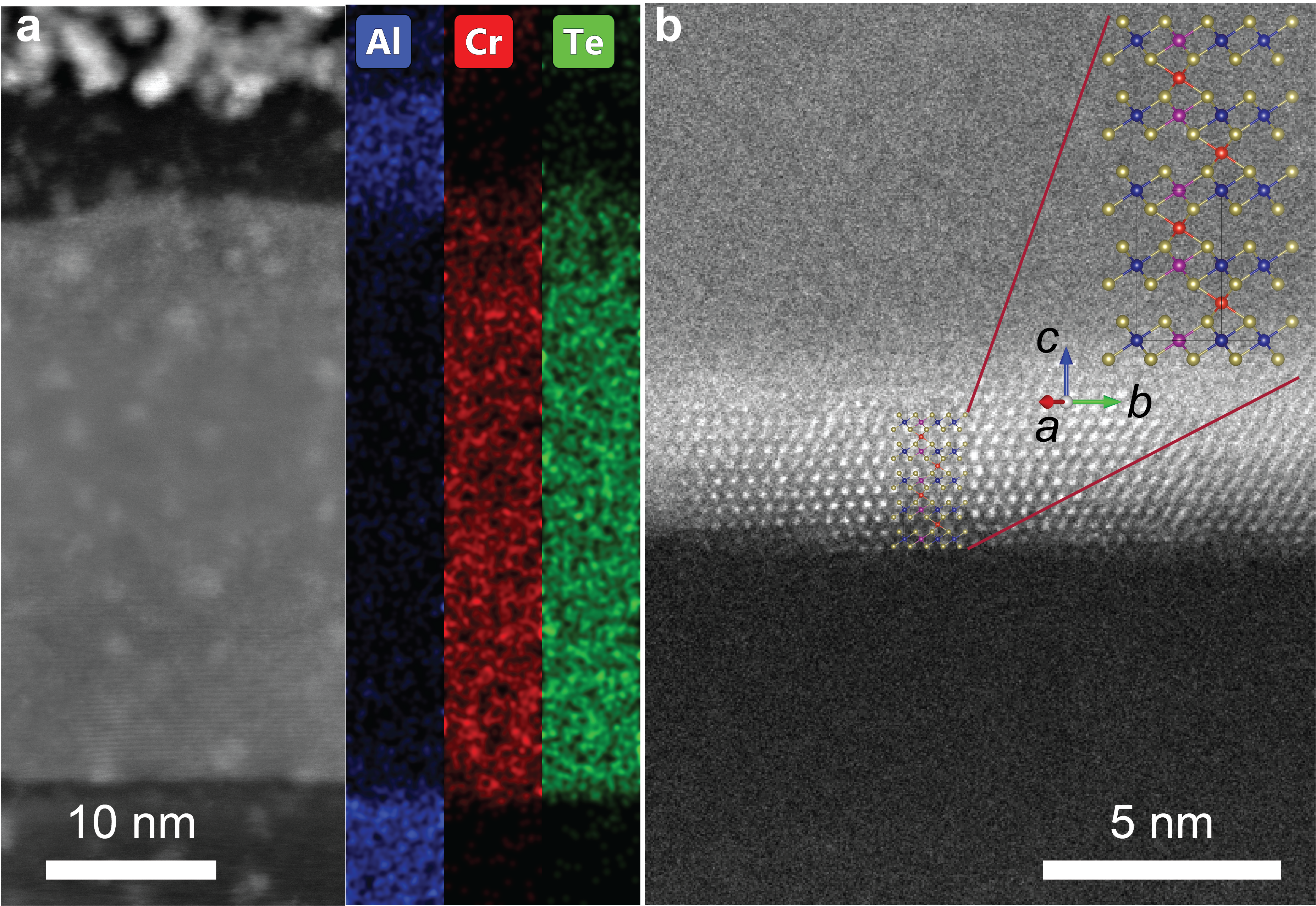}
\mycaption{\label{fig:figs3}
\textbf{Scanning transmission electron microscopy of Cr$_2$Te$_3$.} \textbf{a}, Cross sectional STEM imaging of $c$-oriented Cr$_2$Te$_3$ on Al$_2$O$_3$(0001) with thickness $t$ = 24 u.c.. The energy dispersive X-ray spectroscopy (EDS) profile reveals a uniform elemental distribution. \textbf{b}, HAADF STEM image of a $t$ = 3 u.c. sample illustrating the film quality.}
\end{figure*}

\begin{figure*}%[htb]
\includegraphics{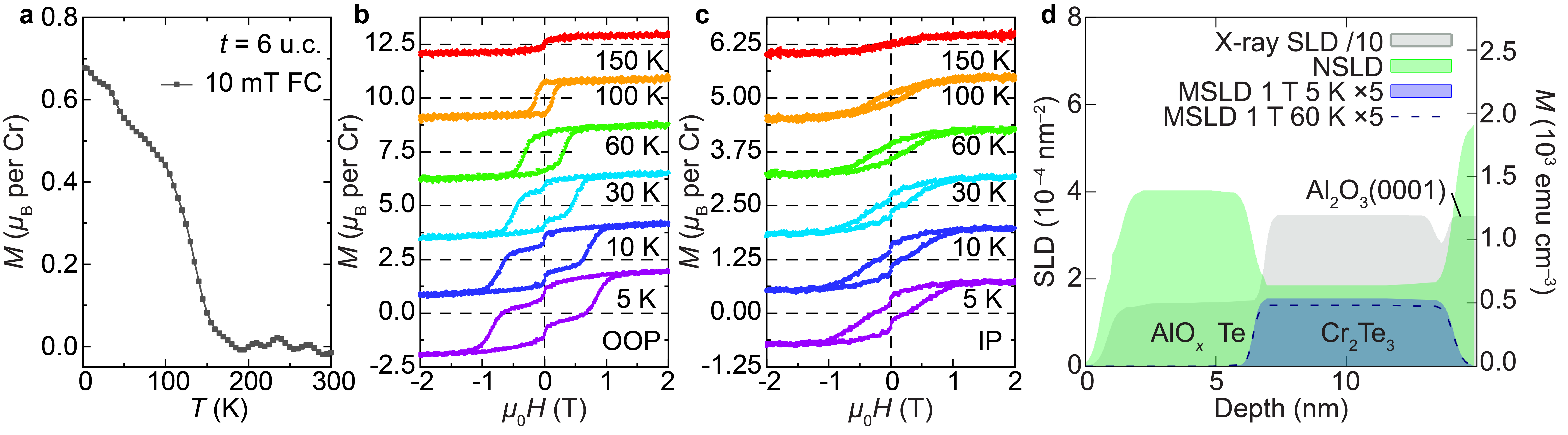}
\mycaption{\label{fig:figs4}
\textbf{Magnetization of 6~u.c.\ Cr$_2$Te$_3$.} \textbf{a}, Temperature dependence of the magnetization $M$ of 6~u.c.\ Cr$_2$Te$_3$ film under the field-cool (FC) condition with an out-of-plane (OOP) magnetic field $\mu_{0}H$ = 10 mT. \textbf{b}-\textbf{c}, Field dependence of $M$ under OOP (\textbf{b}) and in-plane (IP, \textbf{c}) configurations, respectively, at selected $T$. Curves are vertically shifted for clarity. The preference of perpendicular magnetic anisotropy (PMA) is evident in the OOP $M(H)$ scans. \textbf{d}, Depth profiles of polarized neutron reflectometry (PNR) nuclear (NSLD), magnetic (MSLD, with IP $\mu_{0}H$ = 1 T at 5 K and 60 K, respectively) and X-ray scattering length densities (SLD) of 6 u.c. Cr$_2$Te$_3$ on Al$_2$O$_3$(0001) substrate with Te/AlO$_{x}$ capping.}
\end{figure*}

\begin{figure*}%[htb]
\includegraphics{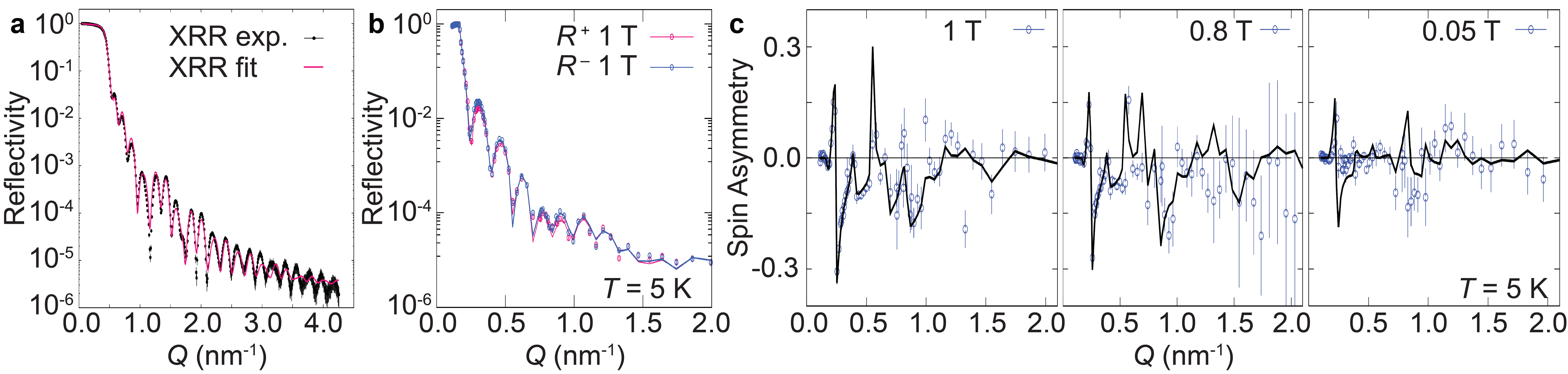}
\mycaption{\label{fig:figs5}
\textbf{X-ray and polarized neutron reflectivity of 24~u.c.\ Cr$_2$Te$_3$ on Al$_2$O$_3$(0001).} \textbf{a}, Measured (points) and fitted (lines) X-ray reflectivity. \textbf{b}, Polarized neutron reflectivity at $T$ = 5 K and $\mu_{0}H$ = 1 T. \textbf{c}, The PNR spin asymmetry ratio SA = $(R^{+} - R^{-})/(R^{+} + R^{-})$ for $\mu_{0}H$ = 1 T, 0.8 T and 0.05 T, respectively. }
\end{figure*}

\begin{figure*}%[htb]
\includegraphics{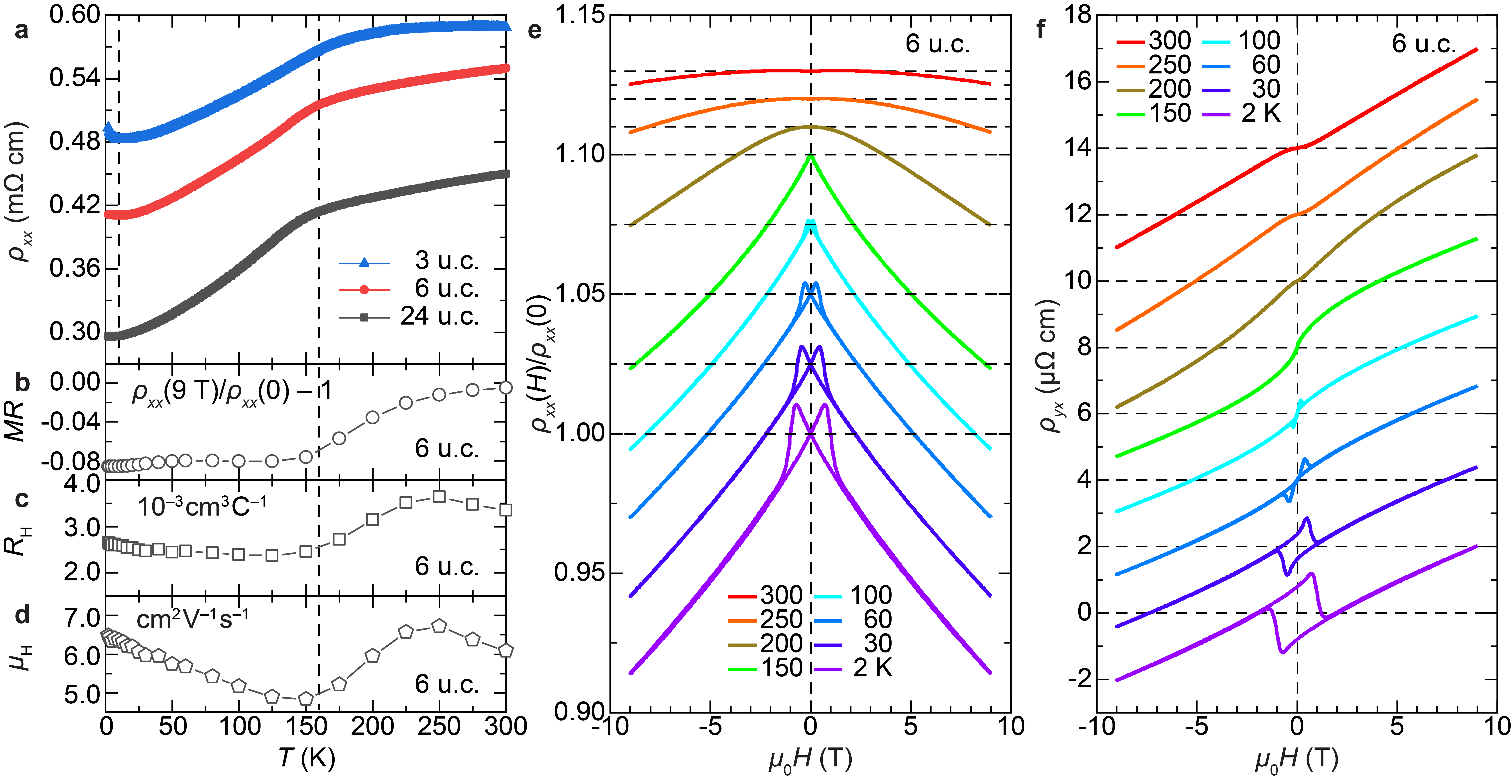}
\mycaption{\label{fig:figs6}
\textbf{Transport properties of Cr$_2$Te$_3$ thin films.} \textbf{a}, Temperature dependence of the longitudinal electrical resistivity $\rho_{xx}(T)$ of Cr$_2$Te$_3$ with $t$ = 3 -- 24~u.c..\ \textbf{b}-\textbf{d}, The key transport parameters for $t$ = 6 u.c., namely, the magnetoresistance [$MR \equiv \rho_{xx}(H)/\rho_{xx}(0) - 1$] at $\mu_{0}H$ = 9 T (\textbf{b}), the Hall coefficient $R\textsubscript{H}$ (\textbf{c}) and the Hall mobility $\mu_{\textrm{H}}$ (\textbf{d}) derived from the linear ordinary Hall effect at 8 -- 9 T. \textbf{e}-\textbf{f}, Magnetic field dependence of $\rho_{xx}(H)$ (\textbf{e}) and the Hall resistivity $\rho_{yx}(H)$ (\textbf{f}) at selected temperatures for $t$ = 6 u.c.. Curves in \textbf{e}-\textbf{f} are shifted vertically for clarity.}
\end{figure*}

\begin{figure*}%[htb]
\includegraphics{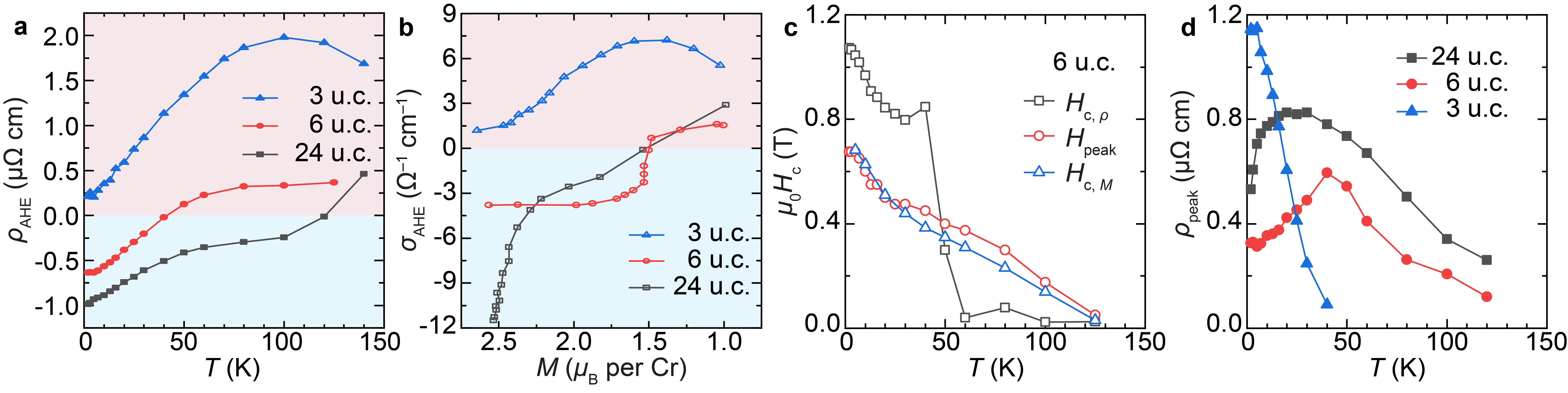}
\mycaption{\label{fig:figs7}
\textbf{The characteristics of the unconventional Hall response of Cr$_2$Te$_3$.} \textbf{a}, Temperature dependence of $\rho\textsubscript{AHE}$ for $t$ = 24 u.c. (black), 6 u.c. (red) and 3 u.c. (blue). \textbf{b}, The corresponding anomalous Hall conductivity $\sigma\textsubscript{AHE}$ as a function of the magnetization $M$. \textbf{c}, Temperature dependence of the field $H\textsubscript{peak}$ at which the hump-shaped peak occurs (red) for $t$ = 6 u.c., along with the coercive fields $H_{\textrm{c,}\rho}$ (black) and $H_{\textrm{c,}M}$ (blue) determined from transport and magnetization measurements, respectively. \textbf{d}, Temperature dependence of the magnitude of the hump-shaped Hall feature $\rho\textsubscript{peak}$.}
\end{figure*}

\begin{figure*}%[htb]
\includegraphics{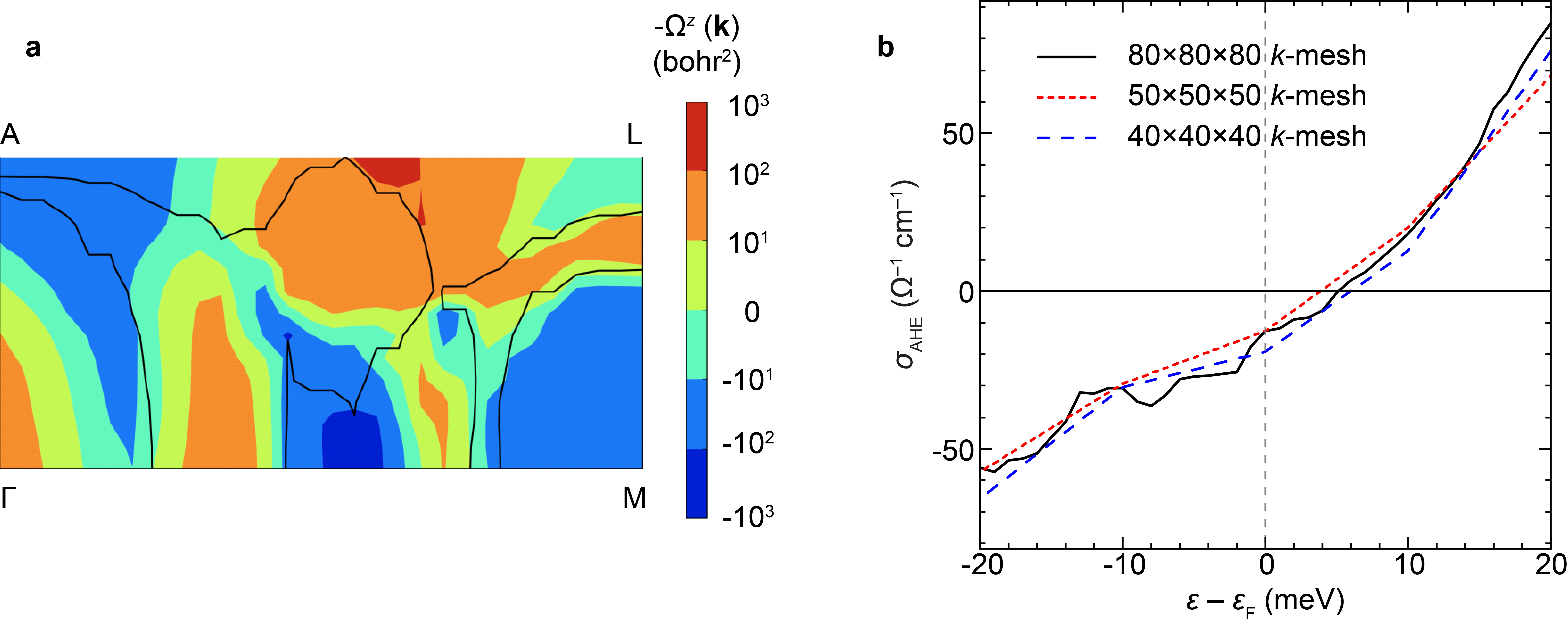}
\mycaption{\label{fig:figs8}
\textbf{Berry curvature and anomalous Hall conductivity in Cr$_2$Te$_3$.} \textbf{a}, Berry curvature contour plot of $-\Omega^{z}(\textbf{k})$ with Fermi surface in the $\Gamma$-M-L-A $k$-plane. \textbf{b}, Convergence test of $k$-mesh for energy dependent $\sigma_{\textrm{AHE}}$ near the Fermi level $\varepsilon_{\textrm{F}}$.}
\end{figure*}

\begin{figure*}%[htb]
\includegraphics{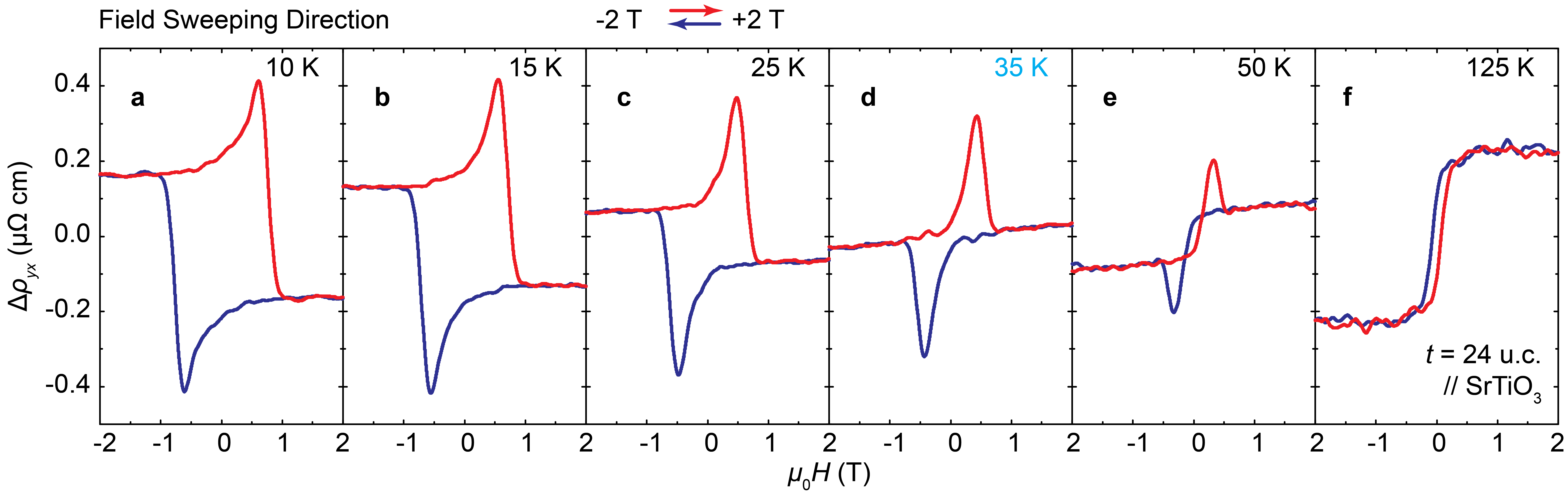}
\mycaption{\label{fig:figs9}
\textbf{Hall response of 24 u.c.\ Cr$_2$Te$_3$ grown on SrTiO$_3$(111).} The Hall traces after removing the linear background at 10 K (\textbf{a}), 15 K (\textbf{b}), 25 K (\textbf{c}), 35 K (\textbf{d}), 50 K (\textbf{e}) and 125 K (\textbf{f}), respectively. Despite the choice of different substrate and hence distinct interface conditions, the temperature dependent anomalous Hall effect sign reversal and the unconventional hump-shaped Hall peaks are also present, attesting to the universality of the observed phenomena.}
\end{figure*}

\nocite{*}

%\clearpage
%\bibliography{1-SI}% Produces the bibliography via BibTeX.